\documentclass[10pt,letterpaper]{article}
\usepackage[top=0.85in,left=1.1in,footskip=0.75in]{geometry}

\usepackage{floatrow}           
\usepackage{graphicx}
\usepackage{dcolumn}
\usepackage{array}
\usepackage{url}
\newcolumntype{P}[1]{>{\centering\arraybackslash}p{#1}} 
\newcolumntype{R}[1]{>{\raggedright\arraybackslash}p{#1}} 

\usepackage{amsmath,amssymb}
\usepackage{changepage}
\usepackage[utf8x]{inputenc}
\usepackage{bbm}

\usepackage{enumitem}

\usepackage[right]{lineno} 
\usepackage{microtype}
\DisableLigatures[f]{encoding = *, family = * }

\newlength\savedwidth

\usepackage[aboveskip=1pt,labelfont=bf,labelsep=period,justification=raggedright,singlelinecheck=off,figurename=Fig]{caption}

\makeatletter
\renewcommand{\@biblabel}[1]{\quad#1.}
\makeatother

\date{}

\usepackage{lastpage,fancyhdr}
\usepackage[]{graphicx} 
\usepackage[outdir=./]{epstopdf}
\usepackage{bm}
\usepackage{color}
\usepackage{bbold}
\usepackage{soul}
\usepackage{epsfig,epic}
\newcommand{\newc}{\newcommand}
\newc{\beq}{\begin{equation}}
\newc{\eeq}{\end{equation}}
\newc{\kt}{\rangle}
\newc{\br}{\langle}
\newc{\beqa}{\begin{eqnarray}}
\newc{\eeqa}{\end{eqnarray}}
\newc{\longra}{\longrightarrow}
\renewcommand{\eqref}[1]{Eq.~(\ref{eq:#1})}

\newcommand{\figref}[1]{Fig.~\ref{fig:#1}}

\newcommand{\tabref}[1]{Tab.~\ref{tab:#1}}

\renewcommand{\phi}{\varphi}

\usepackage{placeins}
\usepackage[numbers,sort&compress]{natbib}

\begin{document}
\vspace*{0.2in}

\begin{flushleft} {\Large \textbf{Is cell segregation like oil and
      water: asymptotic versus transitory regime}\\[1mm]
  }


Florian Franke\textsuperscript{1,2*},
Sebastian Aland\textsuperscript{3,2},
Hans-Joachim B\"ohme\textsuperscript{1,2},
Anja Voss-B\"ohme\textsuperscript{1,2},
Steffen Lange\textsuperscript{1,2}

\bigskip
\textbf{1} DataMedAssist, HTW Dresden, 01069 Dresden, Germany
\\
\textbf{2} Faculty of Informatics/Mathematics, HTW Dresden - University
  of Applied Sciences, 01069 Dresden
\\
\textbf{3} Faculty of Mathematics and Computer Science, TU Freiberg, 09599 Freiberg
\\
\bigskip

* florian.franke{@}htw-dresden.de\\
\end{flushleft}

\section*{Abstract}

Understanding the segregation of cells is crucial to answer questions
about tissue formation in embryos or tumor progression. Steinberg
proposed that separation of cells can be compared to the separation of
two liquids. Such a separation is well described by the Cahn-Hilliard
(CH) equations and the segregation indices exhibit an algebraic decay
with exponent $1/3$ with respect to time. Similar exponents are also
observed in cell-based models. However, the scaling behavior in these
numerical models is usually only examined in the asymptotic regime and
these models have not been directly applied to actual cell segregation
data. In contrast, experimental data also reveals other scaling
exponents and even slow logarithmic scaling laws. These discrepancies
are commonly attributed to the effects of collective motion or
velocity-dependent interactions. By calibrating a 2D cellular automaton
(CA) model which efficiently implements a dynamic variant of the
differential adhesion hypothesis to 2D experimental data from Méhes et
al., we reproduce the biological cell segregation experiments with
just adhesive forces. The segregation in the cellular automaton model follows a
logarithmic scaling initially, which is in contrast to the proposed algebraic
scaling with exponent $1/3$. However, within the less than two orders
of magnitudes in time which are observable in the experiments, a
logarithmic scaling may appear as a pseudo-algebraic scaling. In
particular, we demonstrate that the cellular automaton model can exhibit a range of
exponents $\leq 1/3$ for such a pseudo-algebraic scaling. Moreover,
the time span of the experiment falls into the transitory regime of
the cellular automaton rather than the asymptotic one. We additionally develop a
method for the calibration of the 2D Cahn-Hilliard model and find a match with
experimental data within the transitory regime of the Cahn-Hilliard model
with exponent $1/4$. On the one hand this demonstrates that the transitory
behavior is relevant for the experiment rather than the asymptotic
one. On the other hand this corroborates the ambiguity of the scaling
behavior, when segregation processes can be only observed on short time
spans.

\section*{Author Summary}

Segregation of different cell types is a crucial process for the
pattern formation in tissues, in particular during embryogenesis.
Since the involved cell interactions are complex and difficult to
measure individually in experiments, mathematical modelling plays an
increasingly important role to unravel the mechanisms governing
segregation. The analysis of these theoretical models focuses mainly
on the asymptotic behavior at large times, in a steady regime and for
large numbers of cells. Most famously, cell-segregation models based
on the minimization of the total surface energy, a mechanism also
driving the demixing of immiscible fluids, are known to exhibit
asymptotically a particular algebraic scaling behavior. However, it is
not clear, whether the asymptotic regime of the numerical models is
relevant at the spatio-temporal scales of actual biological processes
and in-vitro experiments. By developing a mapping between 2D cell-based
models and experimental settings, we are able to directly compare
previous experimental data to numerical simulations of cell
segregation quantitatively. We demonstrate that the experiments are
reproduced by the transitory regime of the models rather than the
asymptotic one. Our work puts a new perspective on previous
model-driven conclusions on cell segregation mechanisms.


\section*{Introduction}

Pattern formation of cells and cell segregation are complex and
crucial processes, in particular in the context of embryogenesis. When
different types of cells are intermixed, they start to segregate into
homogeneous domains~\cite{TowHol1955, CerEtal2015, XioEtal2013,
  KayTho2009, CanZarKasFraFag2017}. This behavior has been shown for many different cell
types in several species, for instance hydra~\cite{RieSaw2002, Ste2007}, zebra
fish~\cite{KriEtal2008} and chicken~\cite{MomEtal1995,
FotPflForSte1996}. Why and how cells rearrange themself in a certain
way is still not fully understood, and various theories and hypotheses
have been formulated to explain the process of cell segregation
\cite{KreHei2011, GlaGra1992, GlaGra1993, Ste1970,
  MehVic2013, KriEtal2008, BelThoBruAlmCha2008, NakIsh2011,
  BeaBru2011, MehMonNemVic2012, VosDeu2010, CanZarKasFraFag2017, Har1976}.

One of the most well-known theories in the context of cell segregation
is the differential adhesion
hypothesis of Steinberg~\cite{Ste1970, Ste1963}, which focuses on the
impact of adhesion on cell segregation. He proposed that the sorting
behavior of cells results from differences in the adhesion strengths
between different cell types, which implies that sorting is driven by
the minimization of the surface energy. Additionally, he suggested that a mixed
cell population will always minimize its total adhesive free energy
and conjectured that cells segregate like demixable fluids, e.g.,
water and oil. Note, that this hypothesis is still debated and
alternative, partly related hypotheses where formulated like the
differential surface contraction hypothesis~\cite{Har1976}.

The separation of fluids is theoretically well studied. The kinetics
of this separation can be modeled with the Cahn-Hilliard Navier-Stokes
equations~\cite{LamMau2008, Voo1985, VooGli1988}. The level of
segregation is typically quantified by segregation indices,
the interface length between clusters of different type or by the average
cluster diameter. For a narrow cluster size distribution, the average cluster
diameter scales inverse-proportional to the interface length and segregation
indices, see SI text. An increase of the level of segregation corresponds
to a decrease of the former two measures and, accordingly, an increase
of the latter one, the average cluster diameter. For the Cahn-Hilliard
Navier-Stokes model, it is well known that during segregation the
interface length exhibits an algebraic decay over several orders of
magnitude in time. The exponent of this algebraic scaling depends on
the flows in the model, which is influenced, among others, by the
length scale of the system~\cite{NasNar2017}, ranging from
$1/3$~\cite{WitBacVoi2012, NasNar2017, GarNieRum2003} for the diffusive
regime described by the Lifshitz-Slyozov-Wagner (LSW) theory, into which
the scenario of segregating biological cells falls,
to $2/3$ for the laminar or turbulent regime~\cite{NasNar2017}. Note
that, on the temporal scale, these exponents are only reached
asymptotically and can be
preceded by exponents down to $1/6$ in an intermittent
regime~\cite{GarNieRum2003}. In either case, the average cluster
diameter is inverse-proportional to the interface length, that is both
the cluster diameter and the interface length scale algebraically with
exponents that are equal in absolute value but have opposite signs.

In contrast to fluid segregation, not only one but a variety of
agent-based models have been used to simulate the segregation of
biological
cells~\cite{ZhaThoSwaShiGla2011,MehVic2013,OsbFlePitMaiGav2017}, since
there is a variety of cell-based mechanisms, such as active cells or
cell interaction mechanisms beside adhesion, which have potential
influence on the segregation and need to be studied. While algebraic
scalings of the segregation indices over time can be observed in most
of these models, the corresponding exponents vary over a wide range of
$1/40 - 1/3$, see overview \tabref{scaling_summary}, depending on which segregation mechanisms are
incorporated and which models are used~\cite{BelThoBruAlmCha2008,
  NakIsh2011, BeaBru2011, StrJuuBauKabDuk2014,Dur2021,Kab2012}. One of
the earliest attempts of simulating cell segregation is the
Cellular-Potts-Model (CPM) of Glazier and Graner~\cite{GlaGra1992,
  GlaGra1993}, in which segregation results from differential
adhesion. While the observed segregation indices display a logarithmic
decay, successive studies concluded that the segregation indices
actually follow a logarithmic decay only initially and settle to an
algebraic one for longer
times~\cite{NakIsh2011,BelThoBruAlmCha2008,BeaBru2011,BeaAlmBru2017,Kra2020,Dur2021}.
Nakajima and Ishihara~\cite{NakIsh2011} used the CPM to study the
effects of even and uneven cell type ratios on the segregation
process. They found the exponent of the algebraic scaling to decrease
for increasingly asymmetric mixtures of cells, with exponents ranging
from $1/3$ for a $50/50$ ratio down to $1/4$ for a $90/10$ ratio. In
any case, they observed the average cluster diameter to be
inverse-proportional to the segregation indices. Belmonte et
al.~\cite{BelThoBruAlmCha2008} modeled segregation by a self-propelled
particle model with velocity alignment to study the influence of
collective motion. They also observed algebraic scaling with an
exponent of maximal $0.18$ concluding that even weak collective motion
accelerates cell segregation. Beatrici et al.~\cite{BeaAlmBru2017} used an
active particle approach to compare the segregation behavior under different cellular
interaction mechanisms including that of the DAH but comprising also related principles with
and without collective motion. They measured the average cluster size, which
showed an algebraic decay with exponents ranging from $1/2$, without
collective motion, to $1$, with strong collective motion. The latter
corresponds to exponents between $1/4$ and $1/2$ for the average
cluster diameter. Beatrici and Brunnet~\cite{BeaBru2011}
studied a specific particle system incorporating velocity differences
between cell types, the boids model, and concluded that velocity
differences are sufficient to generate algebraic segregation even
without collective motion. Depending on the chosen velocities and cell
ratios between fast and slow cell types, they observed both logarithmic
and algebraic scaling, the latter with exponents around $1/5$, ranging from $0.18$
to $0.22$. The latter finding is supported by a study of Strandkvist
et al.~\cite{StrJuuBauKabDuk2014} who found an algebraic scaling with
exponents ranging from $0.025$ to $0.17$ with a particle system
incorporating velocity differences between cell types.
Krajnc~\cite{Kra2020} used a vertex model to demonstrate that
differential fluctuations can efficiently sort cells.
He measured the segregation indices over
time, which showed a maximal algebraic decay with exponent of $1/4$.
Durand~\cite{Dur2021} used a CPM with modified update algorithm, which
allows for simulation of larger number of cells over longer times
while preserving cell connectivity. He observed an asymptotic
algebraic decay with an exponent of $1/4$ and concluded that the
previously reported scaling with exponent $1/3$ is only transitory. He
further found the asymptotic scaling to be independent of cell type
ratio and boundary conditions.

Concerning data, several experiments have been conducted on cell
segregation. Rieu and Sawada~\cite{RieSaw2002}, Schötz et al.~\cite{SchEtal2008}
and Beysens et al.~\cite{BeyForGla2000} conducted experiments with
hydra cells and zebra fish cells. They noticed similarities of cell
behavior to fluids
by comparing characteristics of cell segregation with those expected
for viscous fluids according to hydrodynamic laws. For instance, they
compared the ratio of viscosity to surface tension and the time course of
relaxation to the equilibrium and the characteristics of the reached
equilibria. Krieg et al~\cite{KriEtal2008} used gastrulating zebrafish embryos
cells to quantify adhesive and mechanical properties. While doing so, they
also measured the average cluster size over time, which exhibits an
algebraic scaling with exponent $\sim 1/5$, corresponding
to an exponent $1/10$ for the average cluster diameter.
Cochet-Escartin et al.~\cite{CocLocSteCol2017} studied hydra
cells in 3D tissue both in experiments and in CPM simulation to determine whether
differences in tissue surface tension are sufficient for segregation.
They found algebraic scaling with exponent $0.74$ for the experiments
and $0.5$ for the simulations. However, they only measured cell
segregation in the experiments over half an order of magnitude in
time. In contrast, Méhes et al.~\cite{MehMonNemVic2012} studied the influence
of collective motion in experiments with fish and human cells and
measured algebraic cell segregation indices with exponent of $0.31$ for
less than two orders of magnitude in time. They further measured the
average cluster diameter, with an algebraic increase with exponents between
$0.5$ and $0.74$. This means that the cluster diameter was not
inverse-proportional to the segregation indices, indicating that the
cluster size distribution is not narrow. They
suspected that this behavior was a result of collective motion, which
they concluded to be a segregation promoting effect.

In summary, in the context of cell segregation, an algebraic scaling
with an exponent that differs from $1/3$, the value expected for fluid
segregation, has been attributed to additional intercellular
interaction besides differential adhesion~\cite{MehVic2013,
  VisSpaDas2020}. Such mechanisms include collective
motion~\cite{MehMonNemVic2012, FujNakShiSaw2019, BeaAlmBru2017} or velocity-dependent
interaction of the cells~\cite{BelThoBruAlmCha2008,
  BeaBru2011,StrJuuBauKabDuk2014,Kab2012}. The analysis of the asymptotic
behavior in these theoretical models, in a steady regime and for large numbers
of cells, is primarily used to discriminate between models. However,
it is unclear whether this asymptotic regime is relevant for biological
cell segregation processes and the corresponding in-vitro
experiments~\cite{Dur2021}. Moreover, for both experiments and numerical
simulations, the algebraic decay of the segregation indices is usually
only observed during the last two orders of magnitude of
time~\cite{BeaBru2011,NakIsh2011,StrJuuBauKabDuk2014,Kab2012}
or on an even shorter time interval ~\cite{CocLocSteCol2017, MehMonNemVic2012}.

We use an efficient implementation of a 2D cellular automaton (CA) model
according to Voss-B\"ohme and Deutsch~\cite{VosDeu2010}, which solely
incorporates adhesive forces between cells, and develop a direct
mapping between the model parameters and the experimental setup to
reproduce 2D cell segregation experiments from Méhes et
al.~\cite{MehMonNemVic2012}. We find a match between experimental data
and simulations over the whole time span of the experiments. This is
surprising, since our model initially generates logarithmic scaling of
the segregation indices over time, see also
\figref{main_Florian_paper_plot_10_2} in SI. The match between the
model and the proposed algebraic scaling with exponent $1/3$ in the
experiments is possible since the experimental observation is limited
to less than two orders of magnitude in time. To make this point more
pronounced we will use the term pseudo-algebraic scaling for such
behavior in the following. Depending on the model parameters and the
considered time interval, we observe this pseudo-algebraic scaling
with a range of exponents $\leq 1/3$. In the light of such possible
misinterpretations, experimental segregation may actually be explained
solely by adhesive forces between cells. Thus, we propose that, while
additional effects like collective motion might be promoting
segregation, the main factor that governs cell segregation may still be
adhesive forces. Moreover, we also find a match between the
experimental data and the 2D Cahn-Hilliard model. For this comparison, we
develop a mapping between the length scales of the cellular automaton
and the Cahn-Hilliard model, such that only a single parameter of the
Cahn-Hilliard model, the mobility constant which sets the time scale,
has to be fitted. It turns out that the relevant observation window
of the experiments falls in the transitory regime of the Cahn-Hilliard
model, exhibiting an algebraic scaling exponent of $1/4$. Although
Méhes et al.~\cite{MehMonNemVic2012} suggested an algebraic scaling
exponent of $1/3$ for the experimental data, we find a good agreement
with the Cahn-Hilliard model as well, due to the short observation
span. The fact that both models, the cellular automaton and the
Cahn-Hilliard model, both agree with the experimental data, while
exhibiting different scalings for the experimental setup, corroborates
the ambiguity of scaling behavior, when segregation processes are only
observed on short time spans. Even more important, the direct
application to the experimental setup revealed for both models that
the transitory regime of these models is more relevant for the
experimental spatio-temporal scales than the asymptotic regime. Since
biological experiments are by design restricted to finite time spans,
this highlights the importance of considering
additional features of segregation beyond the scaling behavior of
segregation indices, when comparing with theoretical models.

\section*{Results}

\subsection*{Cellular Automaton can reproduce in vitro experiments}

\begin{figure}[ht!]
 \centering

 \includegraphics[scale=0.2]{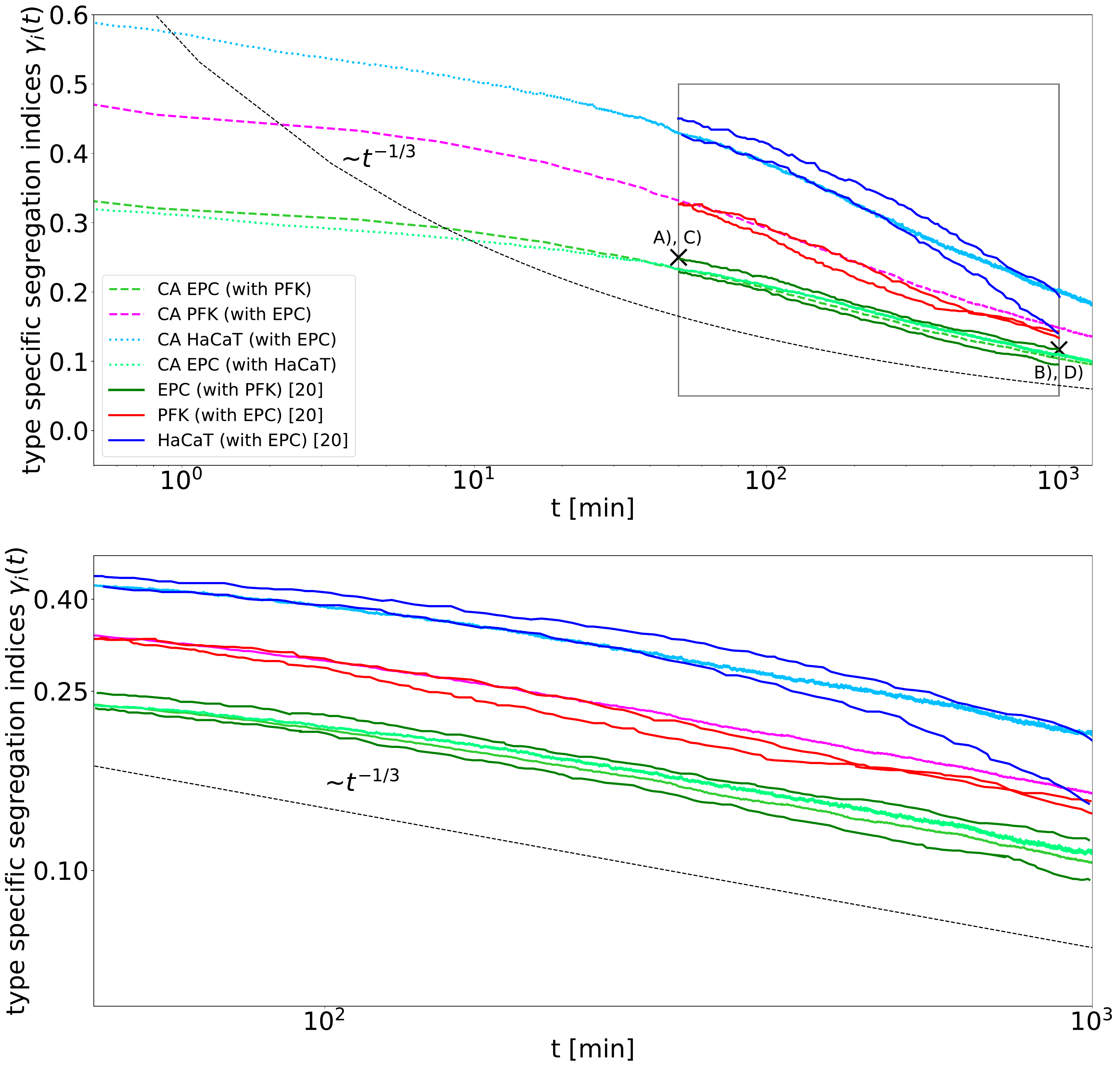}
 \caption{The cellular automaton simulations reproduce the biological
   cell segregation experiments of Méhes et
   al.~\cite{MehMonNemVic2012}. The segregation indices $\gamma_i(t)$
   for the two experiments PFK (dark red) with EPC (dark green) and
   HaCaT (dark blue) with EPC, within the observed time interval
   $50 - 1000$~min, match with the segregation indices predicted by
   the cellular automaton (lines with corresponding brighter colors,
   dashed lines for PFK with EPC and dotted lines for HaCaT with EPC).
   Within the given time interval (grey box in top panel displayed
   again in bottom panel), the segregation indices seem to decay
   algebraically with exponent $1/3$ (black dashed line) as expected
   asymptotically for fluid segregation. For the simulation of the
   segregation indices $\gamma_i(t)$ of PFK ($i=$PFK) mixed with EPC
   ($i=$EPC), we obtain a cell type ratio of
   $N_{\text{PFK}}/N_{\text{EPC}}=41.2/58.8$ and fit the adhesion
   parameters
   $(\beta_\text{PFK-PFK},\beta_\text{EPC-PFK},\beta_\text{EPC-EPC})=(-8.06,-6.56,-0.06)$
   and the time scale of migration
   $\tau_\text{PFK-EPC}\approx 4.2 \text{ min}$. For the simulation of
   the segregation indices $\gamma_i(t)$ of HaCaT ($i=$HaCaT) mixed
   with EPC ($i=$EPC) we obtain a cell type ratio of
   $N_{\text{HaCaT}}/N_{\text{EPC}}=35.2/64.8$ and fit the parameters
   $(\beta_\text{HaCaT-HaCaT},\beta_\text{EPC-HaCaT},\beta_\text{EPC-EPC})=(-7.93,-5.44,0.06)$
   and $\tau_\text{EPC-HaCaT} \approx 35.1 \text{ min}$. In both cases
   $140^2$ cells are simulated, comparable to the cells visible in the
   experiments, starting from a random mixture. Snapshots of the cell
   mixtures at the points marked with crosses labeled A), C) and B),
   D) are displayed in \figref{main_Florian_paper_plot_1_4_3}.}
 \label{fig:main_Florian_paper_plot_1_4_1}
\end{figure}

We compare our cellular automaton simulations with in-vitro data of
Méhes et al.~\cite{MehMonNemVic2012}, see
\figref{main_Florian_paper_plot_1_4_1}. They measured the
segregation indices, cluster sizes, and cluster diameters in the segregation of EPC (fish
keratocyte cell line) with PFK (primary goldfish keratocytes) and
HaCaT (human keratocyte cell line) with EPC over $1.5$ orders of
magnitude in time. The cellular automaton has five parameters, which
are calibrated to the experimental data: Three adhesion parameters
$\text{\boldmath$\beta$}=(\beta_{00}, \beta_{10}, \beta_{11})^T$,
which set the homotypic $(\beta_{11}, \beta_{00})$ and the heterotypic
$(\beta_{01})$ adhesion strengths, the cell type ratio $N_0/N_1$, which
reflects the ratio of all numbers $N_i$ of each cell type in the
segregation experiments, $i\in\{0,1\}$, and the time scale of
migration $\tau$, which relates to the dimensionless time of the cellular
automaton to physical time. While $\tau$ is just a scaling factor for the
time, the segregation indices, that should match between the cellular automaton
and the experiments, are fixed in their ranges and can not be rescaled.
We choose a random initial configuration, which is reasonable with regards to the
experiments which also start with mixed cell configurations, while the
observations commence a bit later. Note that the three adhesion parameters
can be reduced to two effective parameters, the difference of
homotypic adhesion $db$ and the difference between average homotypic
and heterotypic adhesion $\beta^*$, see Materials and Methods for details.
In the experiments, equal areas are covered by each cell type, which
results in different cell numbers due to slightly different cell sizes
for each type. We show that the ratio of cell type numbers $N_0/N_1$
is set by the ratio of the segregation indices
$\gamma_1(t)/\gamma_0(t)$ and thus can be obtained directly from the
experimental data, see \eqref{celltyperatio} and
\figref{main_Florian_paper_plot_7_1}. We check that this ratio
is consistent with the ratio of cell sizes of each type and that the
total numbers of cells of in the experiments and the
simulations are comparable, see Materials and Methods.

Simulations and experiments match well for both cell mixtures, see
\figref{main_Florian_paper_plot_1_4_1}. This match is surprising, as
the cellular automaton displays in the time frame of the experiments rather a logarithmic scaling,
resembling a straight line in the semi-log plot, which contradicts the
proposed algebraic scaling of the data in Méhes et
al.~\cite{MehMonNemVic2012}. However, over just $1.5$ orders of
magnitude in time a logarithmic decay may appear as almost straight
line in a log-log plot as displayed in the bottom panel of
\figref{main_Florian_paper_plot_1_4_1}, making it difficult to
distinguish it from a power law. Therefore, we denote an increase or
decay which approximately follows a straight line in a log-log plot,
but only for a limited time span, as pseudo-algebraic scaling. The
match between the prediction of our model and the experiment in
\figref{main_Florian_paper_plot_1_4_1} demonstrates, that it is not
possible to decide in the limited observation time of experiments
whether the segregation indices decay algebraically or
logarithmically.

For an infinite grid, the asymptotic scaling exponent can only be derived
by theoretical arguments~\cite{WitBacVoi2012, NasNar2017, GarNieRum2003, BerBraHaa2018}.
Since both, our model
system and the experimental system are of finite size, they ultimately
settle at a lower bound of the segregation index, which is dependent on the
system size. However, we choose a sufficiently large system size for the
model, matching that of the experimental setup, such that the lower bound of
the segregation indices is at least one to two orders of magnitude smaller than
the experimentally observed segregation indices. By this, we avoid finite-size
effects and ensure that we can observe the behavior of the segregation indices
in the model even after the observation window of the experiments, such assessing whether the
scaling still changes. In general, it is elaborate to demonstrate that the numerical
behavior of a model is asymptotic. However, for our purpose, it is sufficient
to check whether the scaling changes during or after the observation time to
determine whether it is still transitory. We denote the last measurable scaling
in each simulation as the numerically asymptotic one of the corresponding model, which can
still differ from the theoretically expected value for an infinite-size system.

We observe that for the chosen $\gamma$-fitted parameters the cellular automaton model
reaches its asymptotic regime only below the segregation indices
$\gamma\approx0.15$, exhibiting an algebraic decay with exponent $1/3$
at smaller segregation indices, see
\figref{main_Florian_paper_plot_10_2} in SI for longer simulations. In
contrast, the segregation indices observed in the experiment are
higher ranging from $0.5$ to $0.1$. Therefore, the in-vitro
segregation processes fall into the transitory regime of the
simulations. In this transitory regime, the model, which uses only
adhesive forces, reproduces in-vitro cell segregation. Thus, to
explain the observed scaling behavior, it is not necessary to invoke
additional complex processes or forces which segregate cells, like
collective motion. Note that our main point is to show that the segregation
process is not yet in the asymptotic regime, for which it is sufficient to
demonstrate that the scaling changes during of after the time period where
the segregation indices of the experiment are observed. Eventually, it remains
open whether the last scaling observed numerically in the simulation is actually
the theoretical asymptotic scaling.

In the model, there is a degree of freedom between the time scale
$\tau$ and the adhesion parameters $\beta_{ij}$, see
\eqref{rescaling}. We choose the time scale consistent with the range
of reported average velocities of the cells at low density, which are
$v_\text{PFK}=500~\mu$m/h, $v_\text{EPC}= 30~\mu$m/h, and
$v_\text{HaCaT}=34~\mu$m/h~\cite{MehMonNemVic2012}, such that
$\tau_\text{PFK-EPC} = 2 \Delta x/(v_\text{PFK}+v_\text{EPC}) \approx
4.2$ min and
$\tau_\text{HaCaT-EPC} = 2 \Delta x/(v_\text{HaCaT}+v_\text{EPC})
\approx 35.1$ min with the average length of a cell
$\Delta x \approx \sqrt{350}\mu \text{m}$, see Materials and Methods. With
this choice, the corresponding adhesion parameters are
$(\beta_\text{PFK-PFK},\beta_\text{EPC-PFK},\beta_\text{EPC-EPC})=(-8.06,-6.56,-0.06)$
and
$(\beta_\text{HaCaT-HaCaT},\beta_\text{EPC-HaCaT},\beta_\text{EPC-EPC})=(-7.93,-5.44,0.06)$.
Remarkably, we obtain for both experiments, which we fitted
independently, similar homotypic adhesion parameters for EPC. While
the fitted adhesion parameters may suggest that the homotypic adhesion
of HaCaT and PFK is weaker than that of EPC as well as that the
homotypic adhesion of HaCaT is equal to that of PFK, this fit is not
unique. In fact, due to the short time span the fit is based on, a
wide range of adhesion parameters can reproduce the experimental
observations, as for instance \figref{main_Florian_paper_plot_3_2} below suggests. In
order to refine the fit, additional
data would have to be incorporated, for instance single cell
measurements of adhesion forces of each cell type.

\subsection*{Cahn-Hilliard can reproduce in vitro experiments too}

We also compare the segregation experiments of Méhes et al.~\cite{MehMonNemVic2012}
with fluid segregation. For this, we use the 2D
Cahn-Hilliard model which well describes fluid segregation in the
diffusive regime in terms of a phase-field formulation, see SI text for
details. To fit the parameters of the spatially continuous
Cahn-Hilliard model to the experimental data, which is based on
discrete cells, we develop a mapping between the agent-based cellular
automaton and the Cahn-Hilliard model, see SI text. Due to this mapping,
only the mobility constant $D$ of the Cahn-Hilliard model has to be
fitted to match the time scale of the experiments, while the remaining
parameters can be inferred from the parameters of the cellular
automaton used for \figref{main_Florian_paper_plot_1_4_1}.

\begin{figure}[ht!]
 \centering

 \includegraphics[scale=0.2]{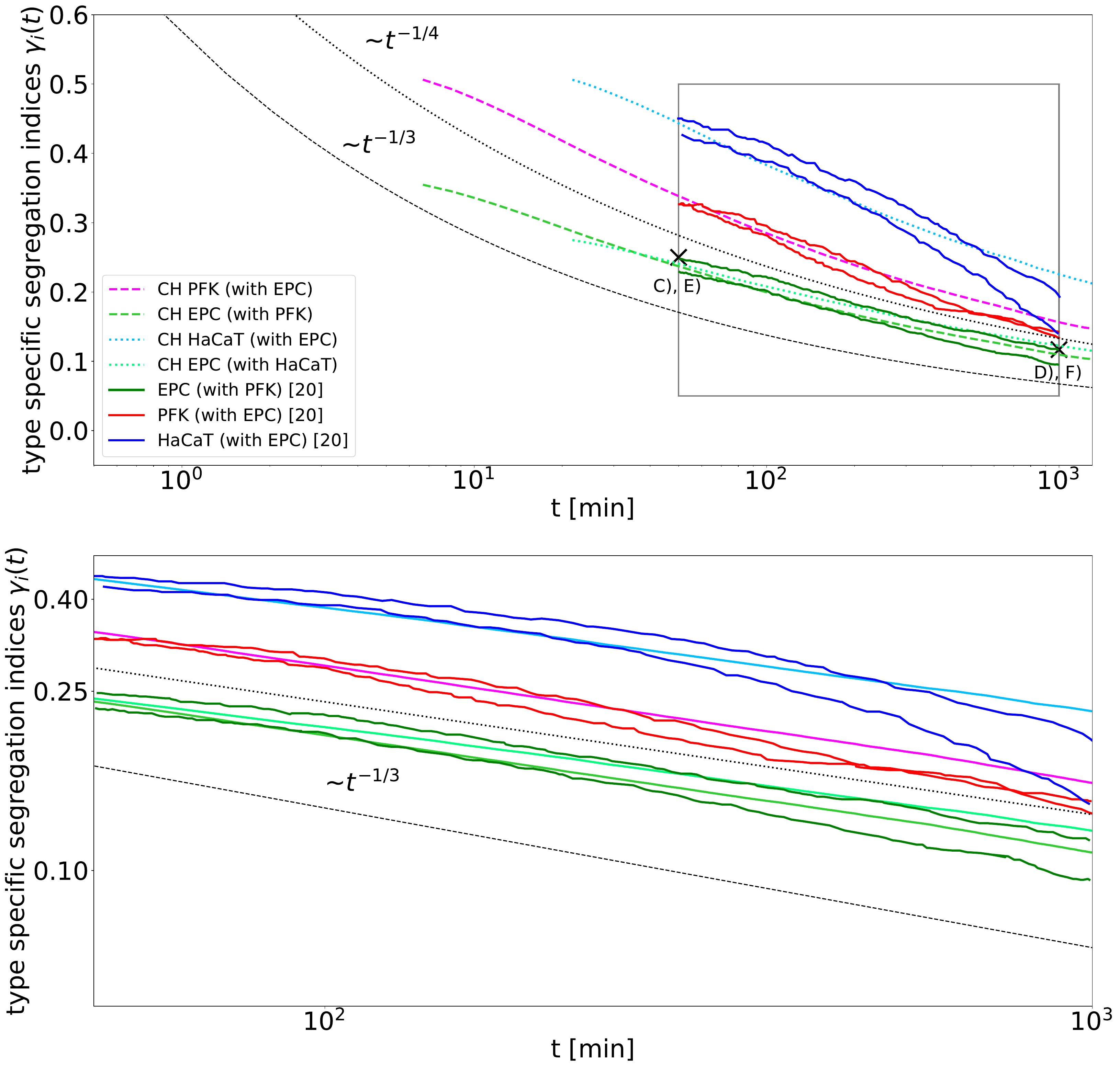}
 \caption{The Cahn-Hilliard simulations reproduce the biological cell
   segregation experiments of Méhes et al.~\cite{MehMonNemVic2012}.
   The segregation indices $\gamma_i(t)$ for the two experiments PFK
   (dark red) with EPC (dark green) and HaCaT (dark blue) with EPC within
   the observed time interval $50 - 1000$~min match the
   segregation indices predicted by the Cahn-Hilliard simulation (lines with
   corresponding brighter colors, dashed lines for PFK with EPC and dotted
   lines for HaCaT with EPC). Within the given time interval (grey box in top
   panel displayed again in bottom panel),
   the segregation indices of the Cahn-Hilliard simulation decay
   algebraically with exponent $1/4$ (black dotted line) rather than
   $1/3$ (black dashed line), which implies that the segregation
   process is in an intermittent regime of fluid segregation, see text
   for details. By using a mapping from the cellular automaton model to
   the Cahn-Hilliard model, see Materials and Methods, parameters
   are set analogous to the parameters used in
   \figref{main_Florian_paper_plot_1_4_1} except for the mobility
   constant $D$, which is fitted to
   $D=36 \text{$\mu $m}^2/\text{min}$ for the mixture of PFK with
   EPC and $D=18 \text{$\mu $m}^2/\text{min}$ for the mixture of
   HaCaT with EPC. Snapshots of the cell mixtures at the points marked
   with crosses labeled C), E) and D), F) are displayed in
   \figref{main_Florian_paper_plot_1_4_3}. Note, that the
   Cahn-Hilliard model is shown after the settling process took place,
   see \figref{main_Florian_paper_plot_9}.}
 \label{fig:main_Florian_paper_plot_1_4_2}
\end{figure}

2D Cahn-Hilliard simulations and experiments match well for both cell
mixtures, see \figref{main_Florian_paper_plot_1_4_2}. The model fits
PFK and EPC very well. However, a small discrepancy can be observed at
the end of the fit of HaCaT from the Cahn-Hilliard model, which
nevertheless reproduces the data as well as the by Méhes et
al.~\cite{MehMonNemVic2012} suggested $1/3$ algebraic scaling exponent.
This match is surprising, as the Cahn-Hilliard simulations rather
display an algebraic decay with exponent of $1/4$ than $1/3$, which
was proposed for the data in Méhes et al.~\cite{MehMonNemVic2012}. However,
within just $1.5$ orders of magnitude in time it is hard to distinguish a
power-law decay with exponent $1/4$ and one with exponent $1/3$.

Note that the segregation indices resulting from the Cahn-Hilliard
model follow only asymptotically $(t\rightarrow\infty)$ an algebraic scaling
with the exponent of $1/3$. This asymptotic decay is usually referred to when
cell segregation is compared to fluid segregation and the exponent
does not depend on the parameters of the Cahn-Hilliard model. However,
the intermittent decay of the segregation indices, before the
asymptotic regime is reached, displays a slower algebraic scaling,
with exponents down to $1/6$~\cite{GarNieRum2003}, and can even
exhibit logarithmic decay, see
\figref{main_Florian_paper_plot_1_4_2}. This intermittent regime
can last for several orders of magnitude in time, and an uneven cell
type ratio can increase the duration of this
regime~\cite{GarNieRum2003}. Furthermore, while the mobility constant
$D$ primarily rescales the physical time in the Cahn-Hilliard model,
we observe that it can also alter the duration
of the intermittent decay. For instance, the simulation
displayed in \figref{main_Florian_paper_plot_9}, which is based
on parameters comparable to the ones used for
\figref{main_Florian_paper_plot_1_4_1} except that the mobility
constant $D$ is several orders of magnitude bigger, already exhibits
an algebraic scaling with the exponent of $1/3$ at segregation indices
$\leq 1/2$.

The determined mobility constants of $D = 36
\text{$\mu $m}^2/\text{min}$ and $18 \text{$\mu $m}^2/\text{min}$ for
PFK with EPC and HaCaT with EPC, respectively, are consistent with the
range of the experimentally measured mobility constants for each cell
type, i.e. PFK ($132 \text{$\mu $m}^2/\text{min}$), EPC
($1.29 \text{$\mu $m}^2/\text{min}$) and, HaCaT ($1.61
\text{$\mu $m}^2/\text{min}$)~\cite{MehMonNemVic2012}. In particular,
the fitted mobility constant for PFK with EPC is greater than that of
HaCaT with EPC, as expected from the individual mobility constants of
each cell type.

\subsection*{Exemplary morphological analysis of both models}

\begin{figure}[ht!]
 \centering
 \includegraphics[scale=0.36]{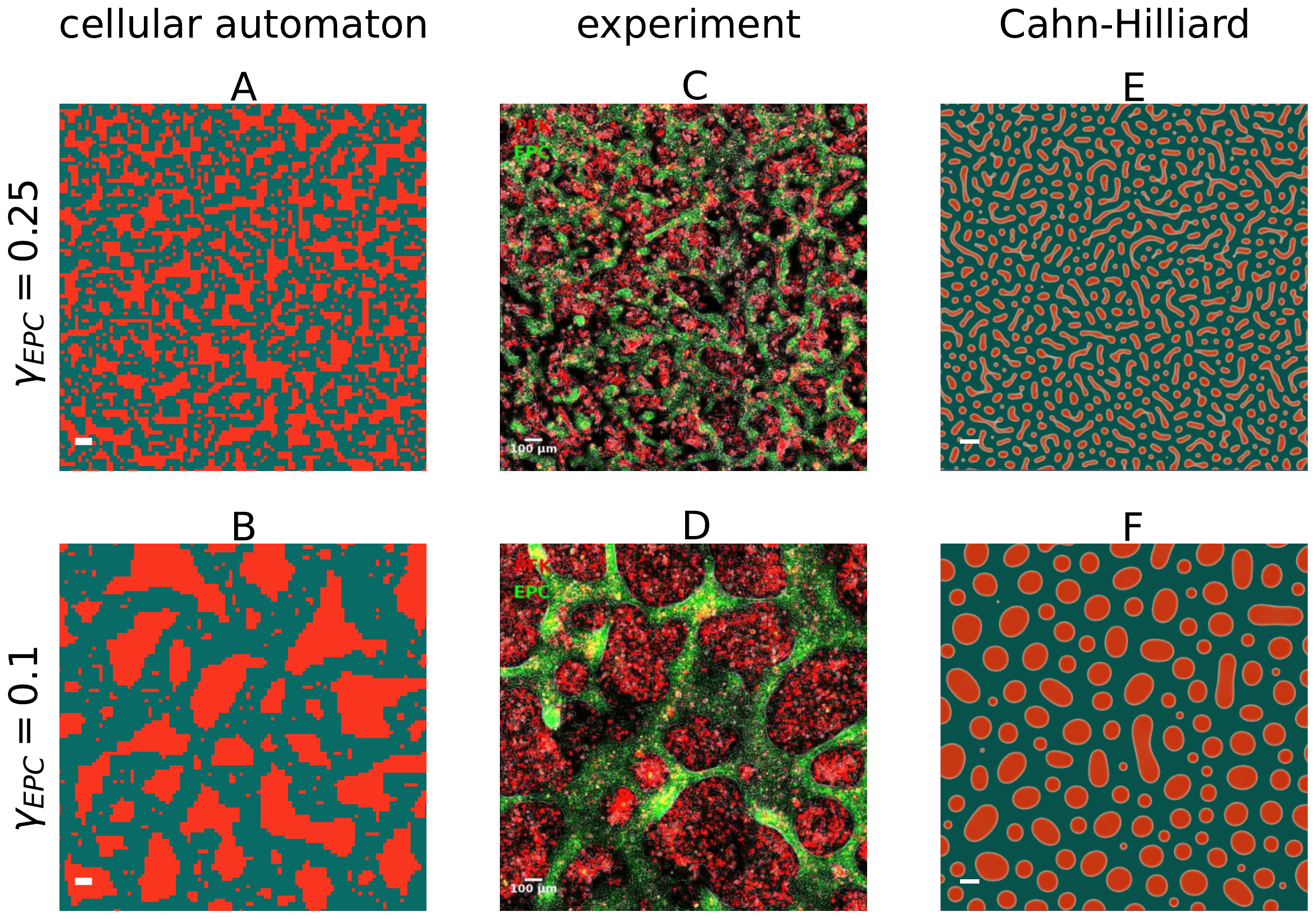}
 \caption{The cellular automaton reproduces morphology and size
   distribution of the cell clusters in the experiments of Méhes et
   al.~\cite{MehMonNemVic2012} of EPC (green) with PFK (red) closer
   than the Cahn-Hilliard model. The snapshots of the cell mixtures
   A), C), E) of the first row are taken at a segregation index of EPC
   $\gamma_\text{EPC}=0.25$, at the start of the experimental
   recording, while the pictures B), D), F) in the second row are at a
   segregation index of EPC $\gamma_{EPC}=0.1$, at the end of the
   recording. A) and B) show the cellular automaton, C) and D) show
   the experiments and are taken from video S5 in Méhes et
   al.~\cite{MehMonNemVic2012}, and E) and F) show the Cahn-Hilliard
   model. The images A), B), E) and F) show a detail from the
   simulations, such that approximately $100^2$ cells are visible, to
   match the spatial scale of the images C) and D) of the experiments.
   The time points corresponding to the images are marked by black
   crosses in \figref{main_Florian_paper_plot_1_4_1} and
   \figref{main_Florian_paper_plot_1_4_2}.}
 \label{fig:main_Florian_paper_plot_1_4_3}
\end{figure}

In conclusion, we observe that two fundamentally different models both
match the experimental segregation indices on the limited time span,
see \figref{main_Florian_paper_plot_1_4_1} and
\figref{main_Florian_paper_plot_1_4_2}. Since the segregation indices
are not sufficient to distinguish between both models with respect to
the experimental observations, we additionally compare the
distribution of cluster sizes $\rho$, the morphology of the clusters, and the
average cluster diameter at two different levels of segregation
qualitatively, see \figref{main_Florian_paper_plot_1_4_3}: In all
three cases, the CA, the CH model and the experiment, the cell type that is less abundant, here PFK shown in
red, forms clusters surrounded by a single contiguous domain of the
more abundant cell type, here EPC shown in green. The Cahn-Hilliard
model results in a rather narrow distribution of cluster sizes while
clusters form circular shapes or slightly elongated bulges, see
\figref{main_Florian_paper_plot_1_4_3}~E) and F). In contrast, the
cells in the experiment of Méhes et al.~\cite{MehMonNemVic2012}
display a wider distribution of cluster sizes with different shapes of
clusters, see \figref{main_Florian_paper_plot_1_4_3}~C) and D).
Interestingly, the configurations of the cellular automaton exhibit
features very similar to the experiment, see
\figref{main_Florian_paper_plot_1_4_3}~A) and B).

\begin{figure}[ht!]
 \centering
 \includegraphics[scale=0.28]{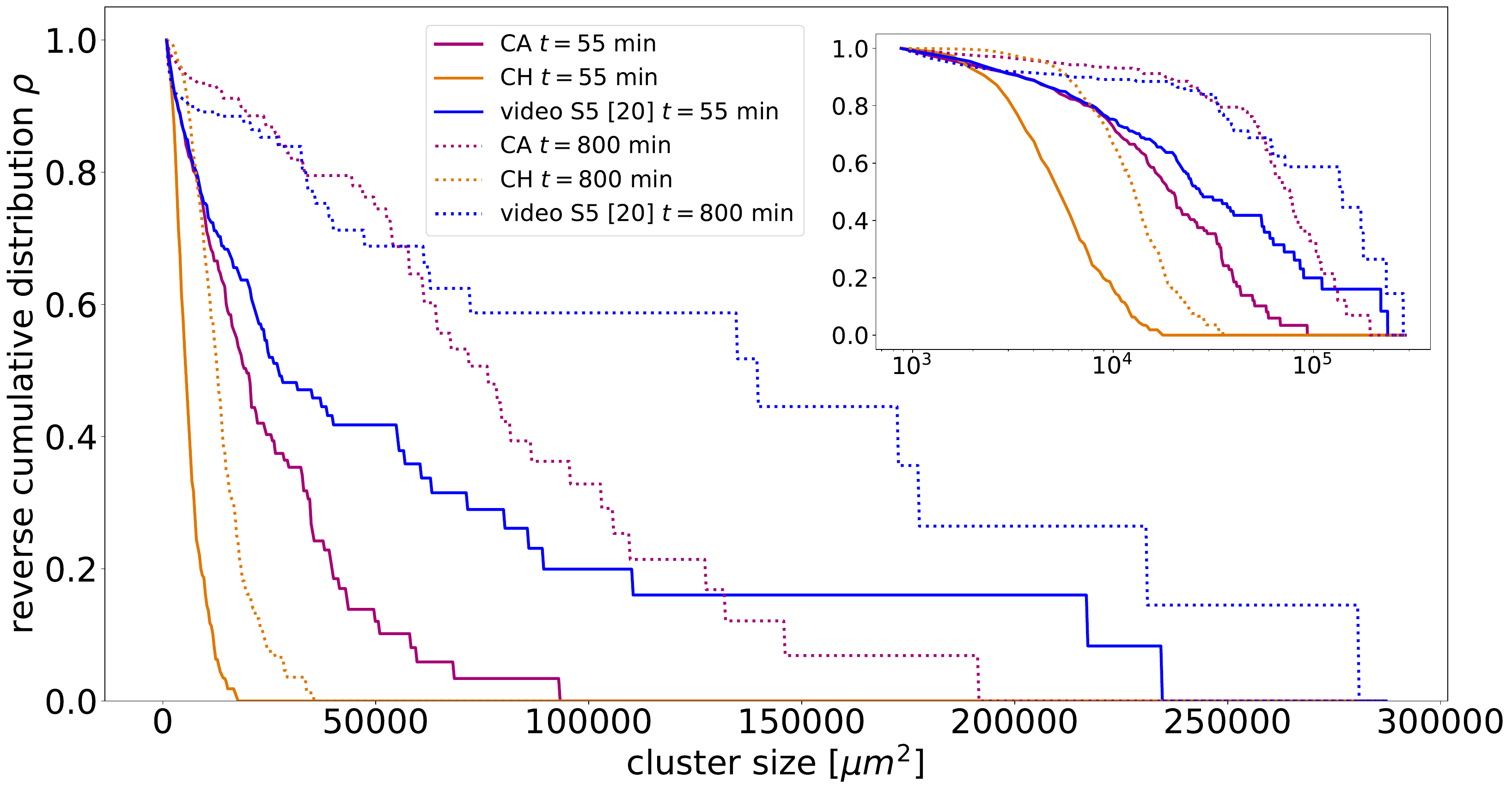}
 \caption{The cellular automaton reproduces the cluster size
   distribution $\rho$ of the experiments of Méhes et
   al.~\cite{MehMonNemVic2012} for EPC with PFK closer than the
   Cahn-Hilliard model. Shown is the
   reverse cumulative probability that a randomly drawn cell belongs to a
   cluster of respective size. For both models and the video S5 from
   Ref.~\cite{MehMonNemVic2012}, two separate cluster size
   distributions are shown, one at an early stage ($t\approx55$min)
   and one at a later stage ($t\approx800$min). The cluster size
   distributions represent exclusively PFK clusters, since EPC as the
   more abundant cell type forms one large connected sea, which we
   ignore in the distributions. Note that clusters below $2$ cells are
   neglected as they can not be resolved in the video, see SI~text.}
 \label{fig:main_Florian_paper_plot_11_3}
\end{figure}

The results of the qualitative comparison of the cell mixtures of
\figref{main_Florian_paper_plot_1_4_3} are confirmed by a quanitative
analysis of the reverse cumulative distribution of cluster sizes $\rho$,
displayed in \figref{main_Florian_paper_plot_11_3}. These
distributions are similar between the cellular automaton and the experiment
for small cluster sizes. Note, that the cellular automaton exhibits an
exponential decay at early times and an algebraic decay with an exponent
$\approx 1$ at later times,
see also \figref{main_Florian_paper_plot_11_2}. In contrast, for the
Cahn-Hilliard model, this distribution declines already steeper at
roughly an order of magnitude smaller cell sizes than for the cellular
automaton model and the experiment. Note that the distribution only
represents the PFK clusters, since EPC cells form a single connected
cluster. The analysis of the experimental data and the computation of the
cluster sizes is detailed in the SI text.

\begin{figure}[ht!]
 \centering
 \includegraphics[scale=0.28]{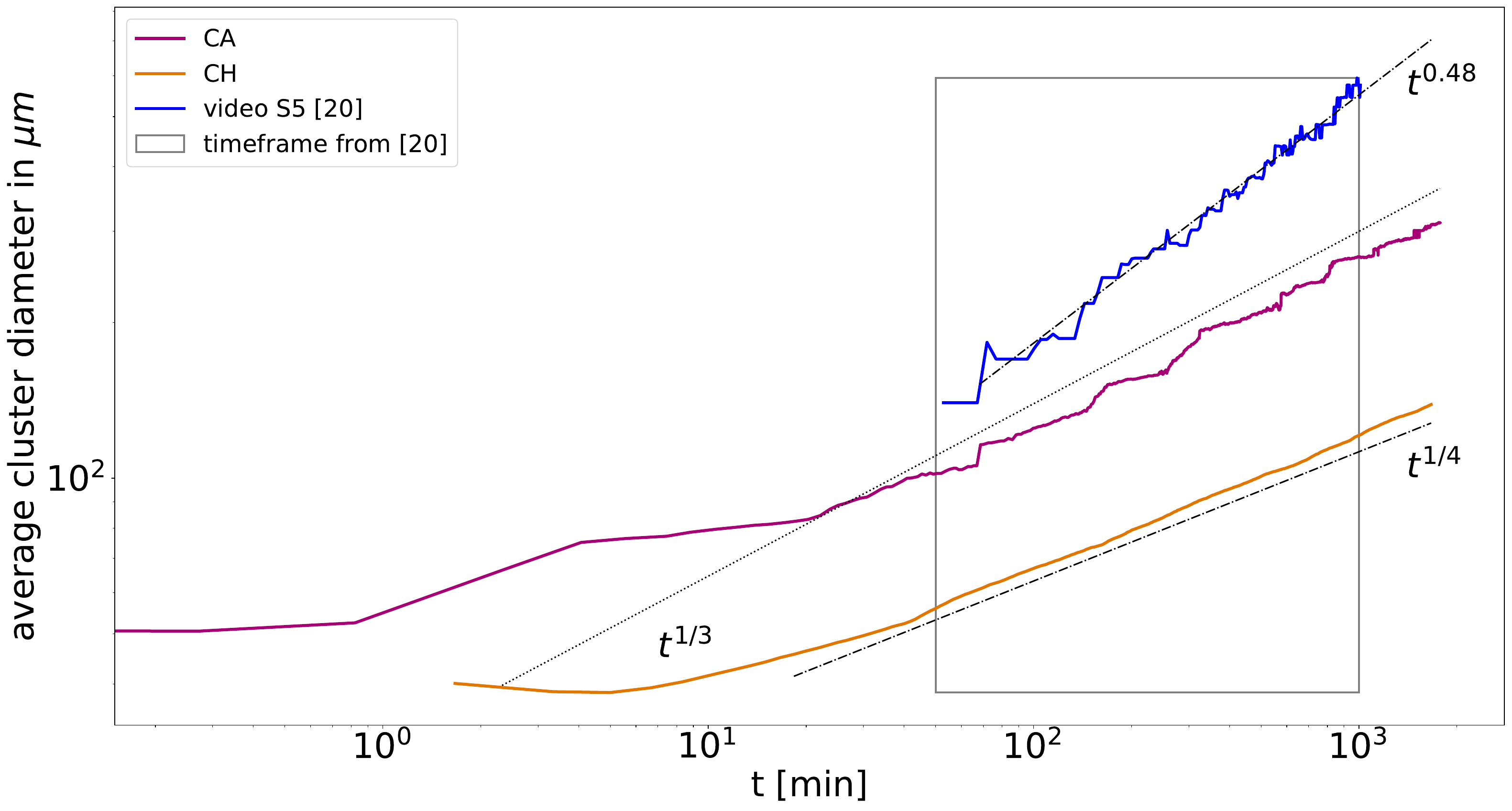}
 \caption{Exemplary comparison of the average cluster diameter in the
   segregation of PFK and EPC for the cellular automaton model (red
   line), the Cahn-Hilliard model (orange line) and the experimental
   data (blue line, based analysis of video S5 of Méhes et
   al.~\cite{MehMonNemVic2012}) computed with two-point correlation
   method, see Materials and Methods. Note that average cluster
   diameter in both models, cellular automaton and Cahn-Hilliard, are
   inverse proportional to their segregation indices. In contrast, the
   average cluster diameter obtained for the experimental data
   displays a steeper power law than expected from the corresponding
   segregation indices.}
 \label{fig:main_Florian_paper_plot_11_1}
\end{figure}

We further use the two point correlation method to obtain the average
cluster diameter. Since Méhes et al.~\cite{MehMonNemVic2012} report
the average cluster diameter of each cell type separately, we
reanalyse the experimentally obtained videos to compute the more
prevalent cluster diameter of both cell types combined. The comparison
of the diameters observed in the models and experiment are displayed
in \figref{main_Florian_paper_plot_11_1}. For the models, we obtain
an average cluster diameter inverse proportional to the
segregation indices with algebraic exponent $1/3$ for the cellular
automaton and $1/4$ for Cahn-Hilliard. In contrast, the experiment
shows an even steeper scaling with an algebraic exponent of $0.48$.

The differences in the length scale of the average cluster diameter are
consistent with the phase images of both models and the experiment,
see \figref{main_Florian_paper_plot_1_4_3}. The Cahn-Hilliard model
displays a very narrow cluster size distribution with more smaller clusters
in comparison to the cellular automaton and the experiment, which display a
much wider distribution with much larger clusters. This results in a
shorter characteristic length scale for the Cahn-Hilliard model. Even though
the cluster size distributions and the cell segregation indices of the cellular
automaton and the experiment are very similar, there are yet significant differences
in the length scale and for the scaling over time of the average cluster diameter.
We attribute this to the differences in cluster shapes. While clusters
appear rounded in the experiment, the clusters in the cellular automaton are still
not rounded. This relates to two competing effects in
cluster formation, growth of the cluster versus rounding of their
interface, and we expect the cluster in the cellular automaton to
become rounder on even longer time scales.

The inverse relation between segregation indices and average cluster
diameter is consistent with previous CPM
models~\cite{NakIsh2011,Dur2021}. In contrast, the steeper increase of
the experimentally observed cluster diameters with exponent $0.48>1/3$
means that the average cluster diameter is not inverse-proportional to
the segregation indices in this case. Méhes et
al.~\cite{MehMonNemVic2012} suspected that this is a consequence of
collective motion, implying that collective motion contributes to a
wider distribution of cluster sizes. Note that, Beatrici et
al.~\cite{BeaAlmBru2017} studied the effect of collective motion in a
segregation model and measured that the algebraic exponent describing
the average cluster size increases with introduction of collective
motion from $1/2$ to $1$ (roughly corresponding to exponents $1/4$ and
$1/2$ for the average cluster diameter). In contrast, the average
cluster size reported by Krieg et al~\cite{KriEtal2008} for the
segregation of gastrulating zebrafish embryos cells display a flatter
power law with exponent of $\approx 1/5$ (roughly corresponding to
exponent $1/10$ for the average cluster diameter).

\begin{table}
  \begin{tabular}[ht!]{c|c|c|c|c}
    experiment & model & $\Delta\gamma[10^{-4}]$ & KSD & KSD \\
    & & & ($t=55$min) & ($t=800$min)\\
    \hline
    PFK and EPC & CA ($\gamma$-fitted) & $0.642$ & $0.3157$ & $0.4660$ \\
    PFK and EPC & CA ($\gamma$-$\rho$-fitted) & $0.725$ & $0.1137$ & $0.2393$\\
    PFK and EPC & CH & $0.939$ & $0.6650$ & $0.8030$\\
    HaCaT and EPC & CA ($\gamma$-fitted) & $1.104$ &  &  \\
    HaCaT and EPC & CH & $2.806$ &  & \\
  \end{tabular}
  \caption{Summary of the averaged mean squared deviation $\Delta\gamma$ and the
  Kolmogorow-Smirnow-Distance (KSD) between each model and the
  corresponding experiment, see Materials and Methods for details.}
  \label{tab:goodness_of_fit}
\end{table}

\subsection*{Exemplary fit optimization for two metrics}

\begin{figure}[ht!]
 \centering
 \hspace*{-2cm}
 \includegraphics[scale=0.4]{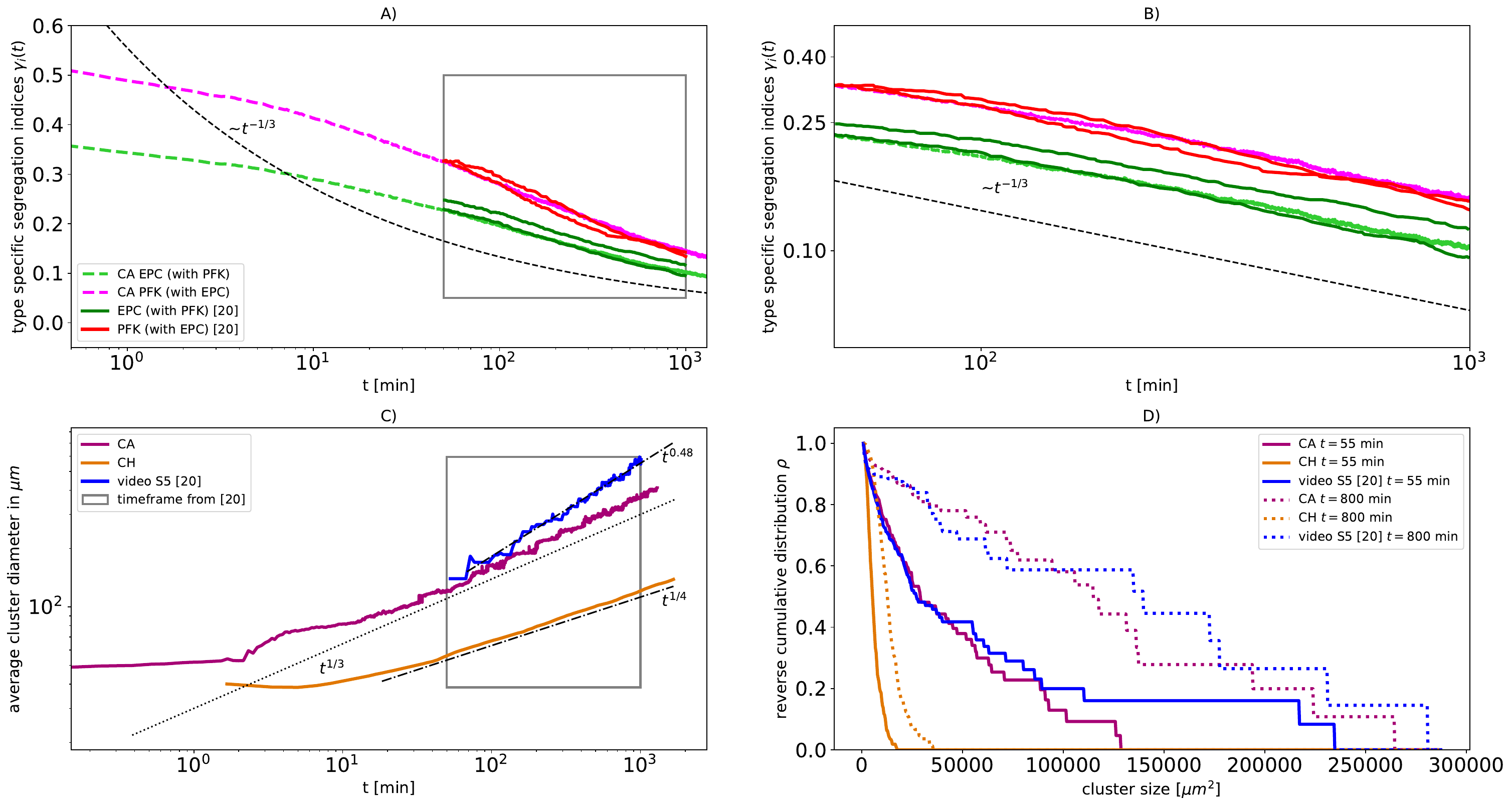}
 \caption{Example representation of the metrics segregation indices $\gamma_i$,
 average cluster diameter and cluster size distribution $\rho$ with $\gamma$-$\rho$-fitted
 parameters for the cellular automaton for the PFK and EPC experiment of Méhes
 et al.~\cite{MehMonNemVic2012}. Figures A) and B) are analogous to
 \figref{main_Florian_paper_plot_1_4_1}, Figure C) is analogous to
 \figref{main_Florian_paper_plot_11_1} and Figure D) is analogous to
 \figref{main_Florian_paper_plot_11_3}. The simulation used $140^2$ cells with
 a cell type ratio of $N_\text{PFK}/N_\text{EPC}=41.2/58.8$, the adhesion parameter
 $(\beta_\text{PFK-PFK},\beta_\text{EPC-PFK},\beta_\text{EPC-EPC})=(-8.0,-5.5,0.0)$
 and a time scale of migration $\tau_\text{PFK-EPC}\approx20.0$ min.}
 \label{fig:main_Florian_paper_plot_13_1}
\end{figure}

The previously presented metrics, average cluster diameter and cluster size
distribution $\rho$, can also be used in the future to improve the fit results of
the model. We have done this exemplary for the cellular automaton and the
experiment PFK and EPC. As indicated before, several parameters can
reproduce the segregation indices similarly well. Thus, the parameter can be
further optimized to fit additional metrics, as demonstrated by an exemplary fit
of both segregation indices $\gamma_i$ and cluster size distribution $\rho$
in \figref{main_Florian_paper_plot_13_1}. As measures for the
goodness-of-fit for the $\gamma$-$\rho$-fitted parameters, we summarize the
averaged mean square deviation $\Delta\gamma$ for the segregation indices,
see Materials and Methods, and the Kolmogorow-Smirnow-distance (KSD) of the cluster
size distributions, in \tabref{goodness_of_fit}.

\begin{figure}[ht!]
 \centering
 \includegraphics[scale=0.3]{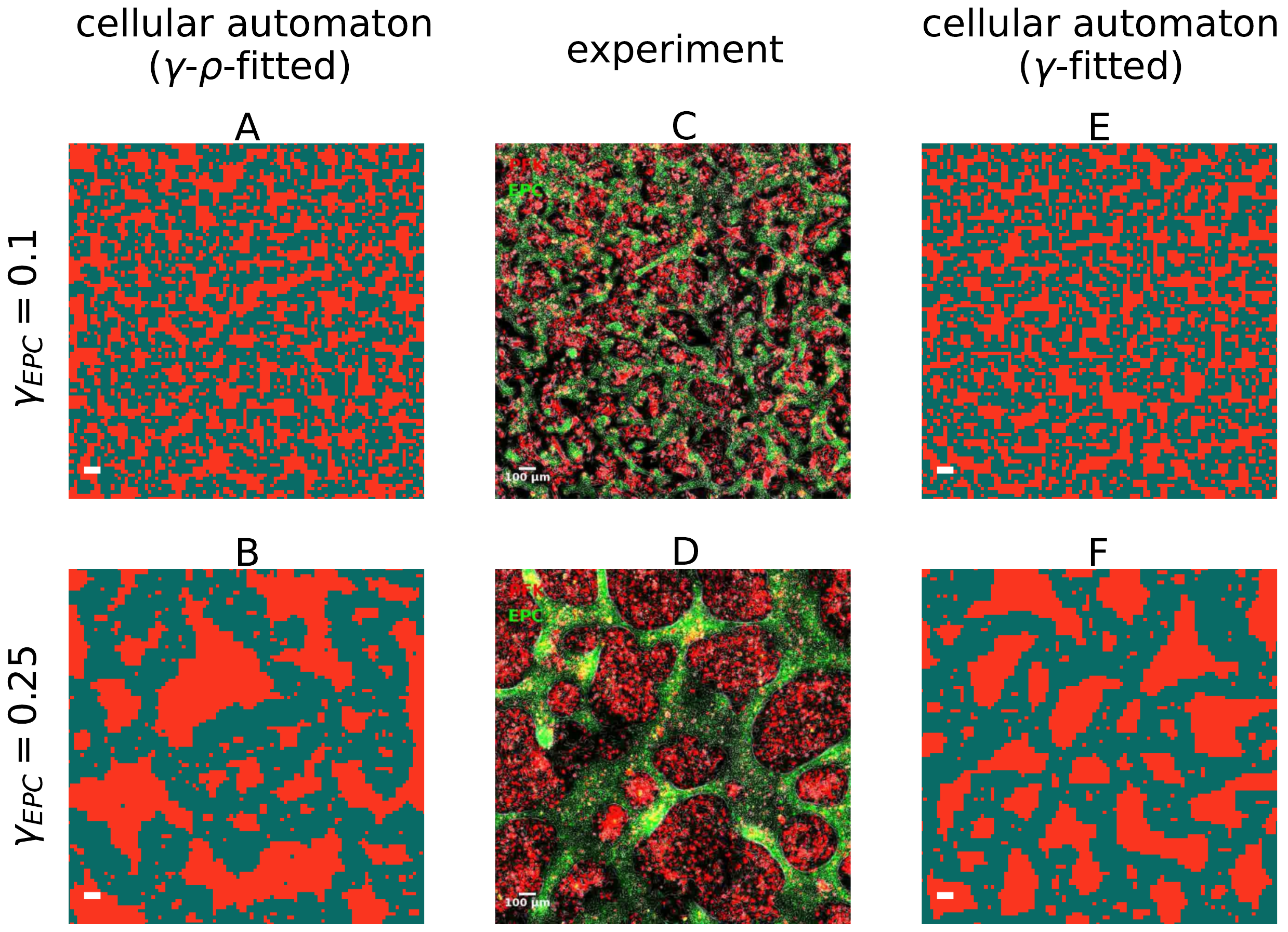}
 \caption{The cellular automaton with $\gamma$-$\rho$-fitted parameters
    reproduces the morphology and size distribution of
    the cell clusters $\rho$ in the experiments of Méhes et
    al.~\cite{MehMonNemVic2012} of EPC (green) with PFK (red) closer than the
    cellular automaton with the $\gamma$-fitted parameters. The snapshots of the
    cell mixtures A), C), E) of the first row are taken at a segregation
    index of EPC $\gamma_\text{EPC}=0.25$, at the start of the experimental
    recording, while the pictures B), D), F) in the second row are at a
    segregation index of EPC $\gamma_{EPC}=0.1$, at the end of the recording.
    A) and B) show the cellular automaton with optimised parameters, C) and D)
    show the experiments and are taken from video S5 in Méhes et
    al.~\cite{MehMonNemVic2012}, and E) and F) show the cellular automaton with
    the $\Delta\gamma$ fitted parameters. The images A), B), E) and F) show a
    detail from the simulations, such that approximately $100^2$ cells are
    visible, to match the spatial scale of the images C) and D) of the
    experiments.}
 \label{fig:main_Florian_paper_plot_13_2}
\end{figure}

The averaged mean square deviation shows, that the cellular automaton reproduces
in any case the experimental segregation indices better than the Cahn-Hilliard
model. Both parameter fits for the CA  reproduce the segregation indices of the
experiment well. Further, the calculated KSD shows that the
$\gamma$-$\rho$-fitted parameters of the cellular automaton reproduce the cluster
sizes of the experiment much better. Exemplary configurations for the two
parameter fits are compared to the experimental observations in
\figref{main_Florian_paper_plot_13_2}.

\subsection*{Parameter influence of the cellular automaton on the segregation}

We have already shown that in the segregating experiments the pseudo-algebraic
scaling can be explained both by the transitory logarithmic scaling
from the cellular automaton and by the transitory algebraic scaling with
exponent of $1/4$ from the Cahn-Hilliard model. Yet, despite the fact
that both models only incorporate adhesion forces, as proposed by
Steinberg, the resulting segregation differs fundamentally between
both models. In addition, in the time frame of the experiment neither
model generates an algebraic scaling with an exponent $1/3$, which is
usually associated with fluid-like segregation. Firstly this
highlights, that not only an algebraic exponent of $1/3$ corresponds
to fluid-like segregation, but exponents between $1/6$ and $1/3$ may
indicate it as well. Secondly, this implies that in contrast to
implicit suggestions of previous works, an exponent differing from
$1/3$ does not necessitate other intercellular interactions or
mechanical forces besides adhesion. In particular, the scaling law
with exponent of $1/3$ only applies to the asymptotic regime of the
models. In contrast, both the Cahn-Hilliard model and the cellular
automaton reproduce the experimental data not in the asymptotic but in
their respective transitory regime, during which the scaling behavior
is more complex and versatile. Additionally this implies, that the
transitory regime of the models has a greater relevance for biological
cell segregation processes than the asymptotic one.

\begin{figure}[ht!]
 \centering
 \includegraphics[scale=0.32]{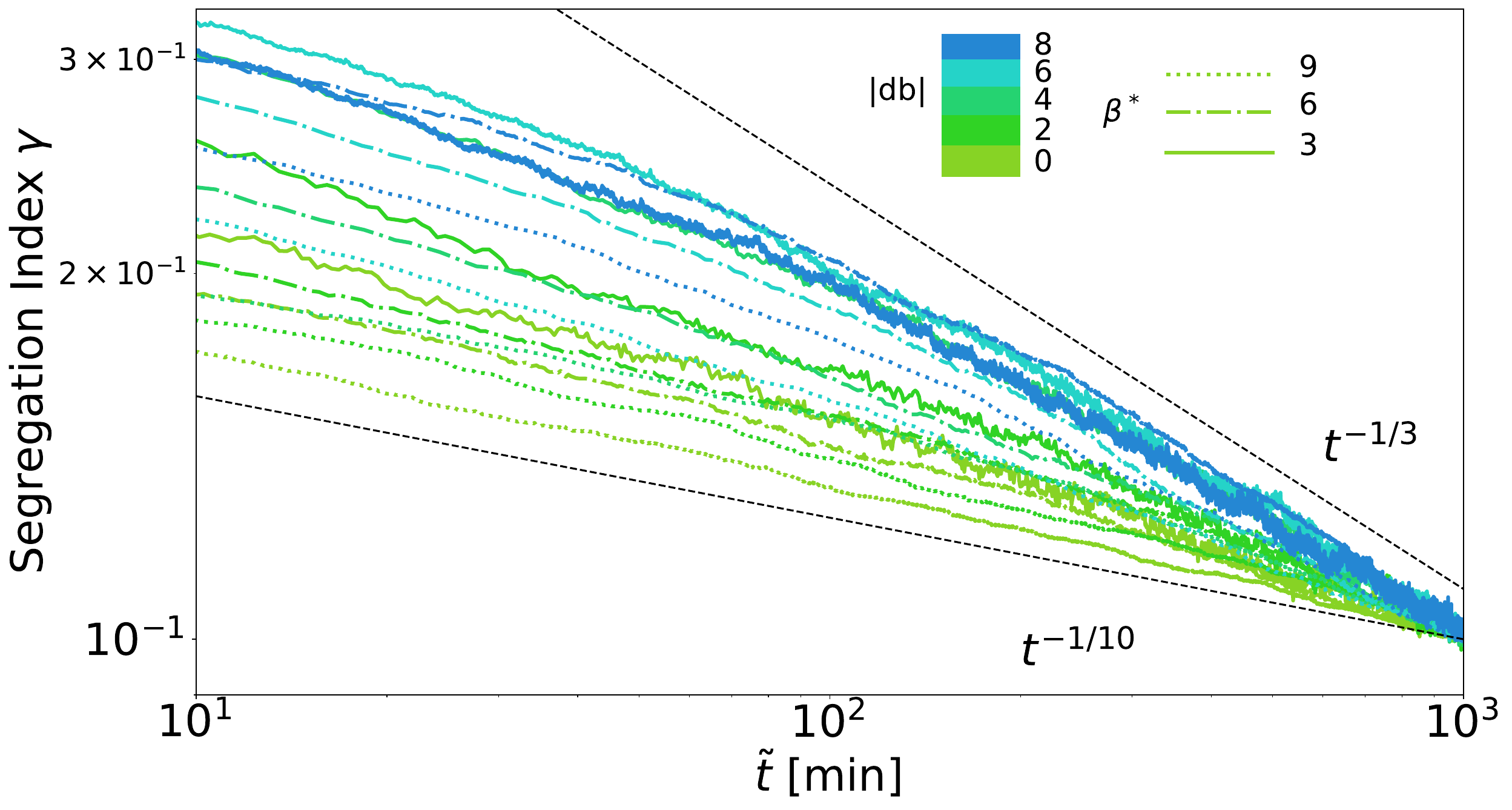}
 \caption{A $1/3$ exponent forms an upper bound for the pseudo-algebraic
 scaling in cellular automaton. Segregation indices obtained
   from the simulation are shown for a range of the effective
   parameters $db$ and $\beta^*$, but only for the last two orders of
   magnitudes in time $\tilde{t}$ before the segregation index reaches
   $\gamma=0.1$ in each simulation. For comparability, the time scale
   of each simulation is set such that all simulations reach
   $\gamma=0.1$ at $\tilde{t}=1$. For each simulation we use $100^2$
   cells, a cell type ratio of 50/50, periodic boundary conditions, and
   a random mixture $\gamma=0.5$ as initial configuration.}
 \label{fig:main_Florian_paper_plot_3_2}
\end{figure}

To relate the range of segregation dynamics displayed by the cellular
automaton to previous experiments, we study numerically the
pseudo-algebraic scaling exponents, which can be generated by the
automaton, and how they depend on the adhesion parameters. The
effective adhesion parameters $db$ and $\beta^*$ determine the
kinetics of the segregation results and therefore, by adjusting these
parameters, we are able to study the impact of those on the possible
exponents. The exact influence of $db$ and $\beta^*$ on the scaling
behavior is complex~\cite{RosBoeLanVos2021}. The cellular automaton is
capable of producing a wide range of pseudo-algebraic scalings, see
\figref{main_Florian_paper_plot_3_2}. The scaling behavior changes
within the experimental regime of segregation indices and is thus
transitory for all displayed parameter choices. Even the flattest
curve close to $t^{-1/10}$ clearly shows this behavior on longer time
scales, see \figref{main_Florian_paper_plot_10_3}. We observe an upper
bound for the exponent of the pseudo-algebraic scaling at $1/3$,
consistent with asymptotic exponents observed in previous particle
models~\cite{NakIsh2011,BelThoBruAlmCha2008,BeaBru2011,
  StrJuuBauKabDuk2014,Kab2012}. Due to the logarithmic decay, the
pseudo-algebraic scaling exponent over two orders of magnitudes
increased with increasing starting time of the observation window,
i.e., it is maximal if the segregation indices at the start of the
observation are small.

\begin{figure}[ht!]
 \centering
 \includegraphics[scale=0.25]{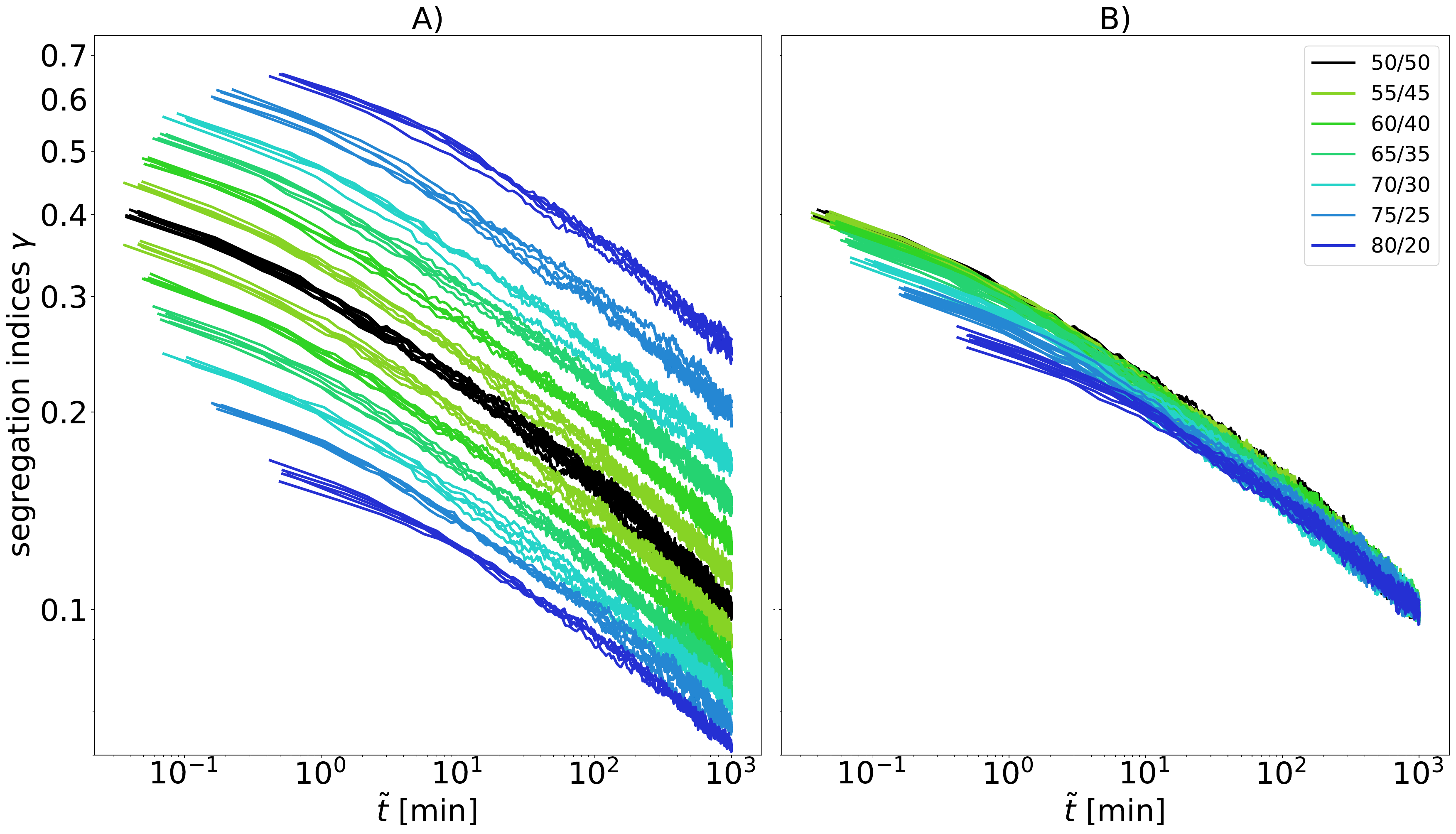}
 \caption{Segregation indices obtained from the simulation shown for
   a range of cell type ratios. For each simulation we use $100^2$
   cells, periodic boundary conditions, $db=0$, $\beta^*=3$, and a
   random mixture as initial configuration. For comparability, the
   time scale of migration $\tau$ of each simulation is set such that all
   simulations reach segregation indices $\gamma_0$ and $\gamma_1$
   with $\gamma_0N_0 = \gamma_1N_1 = 500$ at dimensionless time
   $\tilde{t}=1$. Every color represents a specific cell type ratio,
   while each cell type ratio was simulated five times. A) shows the
   raw data of the simulations. The black lines correspond to an even
   cell type ratio, for which both segregation indices match, while
   for uneven ratios the segregation index of the more abundant cell
   type is below the black line and the other above. B) shows the same
   data where each segregation index $\gamma_i$ is rescaled to a
   segregation index $\tilde{\gamma}_i$ at an even ratio according to
   $\tilde{\gamma}_0 = 2 \gamma_0 N_0/(N_1+N_0)$.}
 \label{fig:main_Florian_paper_plot_3_3}
\end{figure}

However, in contrast to the parameters $db$ and $\beta^*$, the cell
type ratio does not influence the scaling, thus also not the pseudo-algebraic
exponents, which is consistent with recent observations in the CPM
model~\cite{Dur2021}. As shown in Materials and Methods and visualized
in \figref{main_Florian_paper_plot_3_3} A), the cell type ratio just
increases the distance between $\gamma_0$ and $\gamma_1$, but never
the slope in the last orders of magnitudes in time,
\figref{main_Florian_paper_plot_3_3} B).

\section*{Discussion}

We reproduce the experimentally observed segregation indices of Méhes
et al.~\cite{MehMonNemVic2012} by a 2D cellular automaton model which
solely incorporates differential adhesion. The parameters of the model
are calibrated according to the experimental setups. For the
calibration, an efficient algorithm is developed which makes the large
number of simulations required for the exploration of the parameter
space feasible. While Méhes et al. interpreted the decay of the
experimental segregation indices as an algebraic scaling with the
exponent of $1/3$, the cellular automaton model exhibits a logarithmic
decay at the time scale of the experiment, which corresponds to the
transitory regime of the model. We attribute this contradiction to the
limited time span observable in the experiment, which is insufficient
to determine the scaling of the segregation indices. Thus, we refer to
the seemingly algebraic decay observed on a limited time span as
pseudo-algebraic scaling. The match of the experimental results and
the ones from the cellular automaton highlights the possible ambiguity
of scalings on short time spans. We quantify the range of exponents
possible with the pseudo-algebraic scaling of the cellular automaton
model and find the exponent of $1/3$ to be an upper bound. Since an
algebraic decay of the segregation indices with exponent $1/3$ is
commonly considered for fluid-like segregation, and
Steinberg~\cite{Ste1970} proposed that cell segregation is similar to
that of fluids, we additionally compare the experimental results with
fluid segregation expressed by the 2D Cahn-Hilliard model. In order to
adjust the spatial scale of the Cahn-Hilliard model to the cell
segregation experiment, we developed a mapping between the cellular
automaton and the Cahn-Hilliard model. The resulting segregation
indices from the Cahn-Hilliard model fit well the experimental ones,
although they rather follow an pseudo-algebraic decay with exponent $1/4$
than $1/3$ on the relevant time interval, which is again hard to
distinguish on the short time span of the experimental data.

Note that the Cahn-Hilliard model, which well describes fluid
segregation in a diffusive regime, displays an algebraic decay of
segregation indices with exponent $1/3$ only asymptotically. There is
also an intermittent regime, which can last several orders of
magnitude in time, during which exponents down to $1/6$ are
possible~\cite{GarNieRum2003}. The smaller exponent of $1/4$ observed
by us means that for the setup of the experiment the corresponding
fluid-like segregation dynamics are in the intermittent regime. On the
one hand, this highlights that experimentally observed exponents
smaller than $1/3$ do not necessarily rule out fluid like segregation.
On the other hand, this demonstrates the importance of calibrating
segregation models to actual experimental data, as only limited time
regimes of the model may be experimentally relevant.

In conclusion, the calibration of both models, the cellular automaton
and the Cahn-Hilliard model, to the experimental setup reveals that
the transitory regime of these models is relevant for the
spatio-temporal scales of the experiment rather than the asymptotic
regime. This is in contrast to most of the theoretical studies, which
usually focus on the asymptotic regime of the cell segregation models
and do not calibrate the model parameters to the physical constraints
of the experiments. It is reasonable to expect that also for the CPM
the experimental data falls in the transitory regime due to analogies
between segregation processes in the cellular automaton model and the
CPM, especially the analogous structure of the exponent of the cell
switch rates in the cellular automaton and the energy functional in
the CPM. Only recently, asymptotic cell segregation in the CPM has
been explained by directly applying effective adhesion parameters, a
concept previously studied in cellular
automata~\cite{CanZarKasFraFag2017,RosBoeLanVos2021}. Our findings
suggest that future studies on theoretical models and corresponding
numerical simulations of cell segregation should examine not only the
asymptotic regime, but also the complex and less understood kinetics
of the transitory regime.

We present a way to fit both models to experimental data, which can be
applied to future experiments. Since the cell type ratio can directly
obtained from the segregation indices ratio and the time scale of
migration rescales the time scale by a factor, only two parameters
remain to be fitted for the cellular automaton. Note that our
calibration approach should be applicable to other cell-based models,
including the CPM. With respect to the mapping we developed between
the Cahn-Hilliard model and the cellular automaton model, only the
mobility constant $D$ as a single parameter has to be fitted for the
Cahn-Hilliard model.

\subsection*{Issues with scaling analysis}

We point out that in experiments only two or less
orders of magnitudes in time are available to determine the scaling
behavior~\cite{MehMonNemVic2012, KriEtal2008, CocLocSteCol2017}.
Our results suggest, that scaling behavior of segregation indices on
such short time spans is ambiguous, and algebraic scaling on these
time spans should be rather called pseudo-algebraic scaling, since it
may also be a misinterpreted logarithmic decay. This possible
ambiguity has already been hinted at before: Nakajima and Ishihara
mentioned, that their segregation can also be interpreted as a
logarithmic scaling since the algebraic decay was measured only in the
last orders of magnitude~\cite{NakIsh2011}. Belmonte et al. indicated
that a logarithmic decay might be possible, if no coordinated motion
of neighbor cells is present~\cite{BelThoBruAlmCha2008}. We show that
the cellular automaton, which is solely based on differential
adhesion, can generate pseudo-algebraic decays which cover the same
range of exponents $\leq 1/3$ as models which additionally incorporate
other mechanisms like collective motion or differential
velocities~\cite{BelThoBruAlmCha2008, NakIsh2011, BeaBru2011,
  StrJuuBauKabDuk2014}. This wide range of possible segregation
behavior is a feature of the transitory regime while we
observe no steeper scaling than $t^{1/3}$. This puts a
new perspective on conclusions of previous studies, which focused
mainly on the asymptotic behavior of segregation models. In
particular, this implies that deviations of biological segregation
processes from the algebraic scaling with exponent $1/3$ do not rule
out that the segregation is solely based on the minimization of the
total surface energy. In conclusion, due to the ambiguity in the
transitory regime and for short observation spans, it is not possbile
to distinguish between specific models and thus to determine which
mechanisms govern the segregation solely based on the scaling
behavior.

Many studies infer from the scaling behavior of the segregation
indices the impact of certain cell mechanisms, like collective motion on,
cell segregation. In contrast, our results strongly suggest utilizing
additional metrics of segregation when comparing between simulations and
experiments to overcome the ambiguous interpretations of the
segregation indices of experimental data on limited time spans. Such
segregation metrics could be the cluster size distribution $\rho$, the morphology
of the clusters, and the average cluster diameter. As an example of such
an analysis, we compute the cluster size distribution and the average cluster
diameters for PFK with EPC and compare them between models and
experiment. We find that the cellular automaton does not only
reproduce the segregation indices, but also has a more similar cluster
size distribution compared to the experiment, in contrast to the Cahn-Hilliard
model, which misses the large clusters that are present in the
cellular automaton and the experiment. On the other hand, the average
cluster diameter differs between the models and the experiment. For
the models, we obtain, as expected, an average cluster diameter inverse
proportional to the segregation indices with algebraic exponent $1/3$
for the cellular automaton and $1/4$ for Cahn-Hilliard. In contrast,
the experiments display a steeper algebraic scaling with exponent
$0.48$, meaning that the average cluster diameter is not inverse
proportional to the segregation indices, which has been attributed to
collective motion~\cite{MehMonNemVic2012}. In conclusion, the cellular
automaton reproduces the experimental cell segregation better than the
Cahn-Hilliard model, but still misses features which may be related to
collective motion, but are not incorporated in the model. In fact,
the similarities between experiment and cellular automaton in the
cluster size distribution $\rho$ and the segregation indices $\gamma_i$ together with
the differences in the average cluster diameter point towards
differences in cluster shapes between model and experiment.
Note that all models display scaling behavior consistent with the
experimental one. Only by considering additional metrics, in our case
the cluster size distribution, and directly comparing the corresponding
time series between experiment and the calibrated models, a distinction
between the models becomes possible.

Note, that the segregation in the cellular automaton follows the
diffusion-and-coalescence mechanism~\cite{BerBraHaa2018}. In
particular, the diffusion of clusters is driven by fluctuations of
cells at the clusters' boundaries, see also the exemplary video
S1-Movie in SI. The diffusion-and-coalescence mechanism is usually
associated asymptotically with an algebraic scaling with exponent
$1/4$~\cite{BerBraHaa2018, Dur2021, Kra2020, BeaAlmBru2017,
  NakIsh2011}, reflecting the competition between the two effects
driving segregation: the growth of clusters versus the rounding of
their interfaces. However, at the intermediate time scales considered
here the clusters in the cellular automaton are not sufficiently
rounded yet, which most likely causes a steeper scaling with exponent
$1/3$.

\subsection*{Additional observations}

Note that the inverse relation between segregation indices and average
cluster diameter is consistent with observations in previous CPM
models~\cite{NakIsh2011,Dur2021}. In addition the range of exponents observed
in our cellular automaton model is consistent with previous models of 2D cell
segregation without collective
motion~\cite{NakIsh2011,BeaBru2011,BeaAlmBru2017,Kra2020,Dur2021},
while the addition of collective motion accelerates segregation
leading to larger exponents~\cite{BelThoBruAlmCha2008,BeaAlmBru2017}.
The biggest difference between the cellular automaton and the
experiment is the steeper increase of the cluster diameter in the experiment with
exponent $0.48$. Note, however, that cluster sizes reported by Krieg et
al~\cite{KriEtal2008} for the segregation of gastrulating zebrafish
embryos cells display a much flatter power law with exponent of
$\approx 1/5$ (roughly corresponding to exponent $1/10$ for the
average cluster diameter).

Another interesting feature of cell segregation is which cell type
encloses the other. While it seems reasonable that the more abundant
cell type should enclose the other, Beatrici and
Brunnet~\cite{BeaBru2011} have found different behavior depending on
the cell type ratios and the cells' velocities. In addition, to
resolve the contradicting logarithmic decay found by Glazier and
Graner~\cite{GlaGra1993} in the CPM and the algebraic scalings found
in successive studies with CPM~\cite{NakIsh2011} and particle
models~\cite{BelThoBruAlmCha2008,BeaBru2011}, Nakajima and
Ishihara~\cite{NakIsh2011} proposed that the number of cells
considered in a simulation affects the scaling behavior.
Our results suggest that rather the time regime determines the
scaling exponents observed over one or two orders of magnitudes.
This is consistent with the fact that many simulations display a logarithmic
decay initially, independently of the number of
cells~\cite{BeaBru2011, BelThoBruAlmCha2008, CocLocSteCol2017, NakIsh2011}.
This is further supported by Beatrici and
Brunnet~\cite{BeaBru2011}, which found no difference in the scaling
behavior for a wide range of cell numbers ($500$ to $8000$) in their
simulations. Recently Durand~\cite{Dur2021} also questioned the effect
of the numbers of cells on the scaling behavior. Likewise, we observe the same
logarithmic decay for a range of $25^2$ to $140^2$ cells per simulation, while
only the fluctuations of the segregation indices are diminished by using more
cells.

\section*{Materials and Methods}

\subsection*{Cellular automaton: model and calibration}

For simulating cell segregation, we use a cellular automaton based on
Voss-B\"ohme et al.~\cite{VosDeu2010}. We use a 2D-quadratic lattice
$S$ with $N\in\mathbb{N}$ nodes for each dimension,
$S=\{1,...,N\}\times \{1,...,N\}$. We assign exactly one cell to each
node. Each cell has the area of $(\Delta x)^2$, which leads to lattice
lengths $N\Delta x$ for each side. Every cell is mapped to a specific
cell type $W=\{0,1\}$ with $\xi:S\rightarrow W$ defining a specific
configuration of cells on the lattice. Based on two possible cell
types, we define three adhesion parameters
$\text{\boldmath$\beta$}=(\beta_{11}, \beta_{10}, \beta_{00})^T$ which
set the stickiness of two directly neighboring cells depending of
their type. The more two neighboring cells stick to each other, the
larger the associated $\beta_{ij}$ parameter. $\tau$ denotes a
parameter to adjust the time scale of migration in the simulation.
Further, based on these parameters, the rate $r(\textbf{x},\textbf{y})$ of two
cells at neighboring positions
$\textbf{x},\textbf{y} \in S, |\textbf{x}-\textbf{y}|=\Delta x$
swapping their locations is given by:
\begin{equation}
  \label{eq:rate1}
  r(\textbf{x},\textbf{y},\xi)=
    \begin{cases}
      \tau^{-1}\exp\{\beta_\text{sum}(\textbf{x},\textbf{y},\xi)\}&\text{, if } \xi(\textbf{y}) \neq \xi(\textbf{x}) \text{ and } |\textbf{x}-\textbf{y}|=\Delta x \\
      0&\text{, otherwise}
    \end{cases}
\end{equation}
where
\begin{equation}
  \label{eq:rate2}
    \beta_\text{sum}(\textbf{x},\textbf{y},\xi)=-\sum_{\textbf{z}:|\textbf{z}-\textbf{x}|=\Delta x}^{} \beta_{\xi(\textbf{x})\xi(\textbf{z})} - \sum_{\textbf{z}:|\textbf{z}-\textbf{y}|=\Delta x}^{} \beta_{\xi(\textbf{y})\xi(\textbf{z})}\text{.}
\end{equation}
Notice that the definition of the homotypic adhesion parameters in
~\eqref{rate1} and ~\eqref{rate2} is such that smaller (or more negative) parameters lead
to higher migration rates and therefore represent lower adhesion forces.
Instead of using the usual Metropolis algorithm and Monte-Carlo steps,
this model is implemented in continuous time by applying the idea of
the Gillespie algorithm to the cellular automaton, see SI text for
details, which results in a speed-up of the simulations by several
orders of magnitude.

Further, the cellular automaton simulates segregation and thus the
segregation indices $\gamma(\tilde{t})$ in a dimensionless time
$\tilde{t}$. The time scale of migration $\tau$ which transforms this
dimensionless time $\tilde{t}$ into physical time $t$, as
$t = \tau\tilde{t}$, is calibrated based
on the experimental data. By matching the
physical time $t_{\gamma=\gamma_\text{match}}$ at which the
experimental segregation indices first reaches a particular value
$\gamma_\text{match}$, such that
$\gamma_\text{exp}(t_{\gamma=\gamma_\text{match}})=\gamma_\text{match}$
and the dimensionless time $\tilde{t}_{\gamma=\gamma_\text{match}}$ at
which the simulated segregation indices first reaches this value
$\gamma_\text{sim}(\tilde{t}_{\gamma=\gamma_\text{match}})=\gamma_\text{match}$,
and estimate
$\tau = t_{\gamma=\gamma_\text{match}} /
\tilde{t}_{\gamma=\gamma_\text{match}}$. If no value for $\tau$ is
provided, it is set to $1$ dimensionless and therefore neglected.

Voss-B\"ohme et al.~\cite{VosDeu2010} proposed an effective parameter $\beta^*$
for two cell types, which determines the asymptotic sorting behavior, where
\begin{equation}
  \beta^*=\beta_{00}+\beta_{11} - 2\beta_{10}\text{.}
\end{equation}

The impact of this parameter has been numerically confirmed and generalized to
an arbitrary number of cell types by Rossbach et al.~\cite{RosBoeLanVos2021}.
We reparametrize the adhesion parameters $\text{\boldmath$\beta$}$
based on the effective parameter $\beta^*$ to better describe the
impact of the parameters on the segregation behavior:
\begin{equation}
  db=\beta_{11}-\beta_{00}\text{,}
\end{equation}
\begin{equation}
  d=\beta_{00}+\beta_{10}+\beta_{11}\text{.}
\end{equation}
This leads to the following invertible transformation equation:
\begin{equation}
  \text{\boldmath$\beta$}=\begin{pmatrix} \beta_{00} \\ \beta_{10} \\ \beta_{11} \end{pmatrix} = \frac{\beta^*}{3} \begin{pmatrix} \frac{1}{2} \\ -1 \\ \frac{1}{2} \end{pmatrix} + \frac{d}{3} \begin{pmatrix} 1 \\ 1 \\ 1 \end{pmatrix} + \frac{db}{2} \begin{pmatrix} -1 \\ 0 \\ 1 \end{pmatrix}\text{.}
\end{equation}
The parameter $d$ rescales the rates in a trivial way, since an
increase of $d$ by $\Delta d$ will increase all $\beta_{ij}$ by the
same amount $1/3\Delta d$ and therefore decrease all rates by a factor
$\exp\{-8/3\Delta d\}$, independently of $\xi,x,y$:
\begin{equation}
  \label{eq:rescaling}
  \begin{split}
    r(\textbf{x}, \textbf{y}) & = \tau^{-1} \exp\left\{-\sum_{\textbf{z}:|\textbf{z}-\textbf{x}|=\Delta x}{\left(\beta_{\xi(\textbf{x})\xi(\textbf{z})}+\frac{1}{3}\Delta d\right)} -\sum_{\textbf{z}:|\textbf{z}-\textbf{y}|=\Delta x}{\left(\beta_{\xi(\textbf{y})\xi(\textbf{z})} +\frac{1}{3}\Delta d\right)}\right\}\\
    & = \tau^{-1} \exp\left\{-\frac{8}{3}\Delta d\right\} \exp\left\{-\sum_{\textbf{z}:|\textbf{z}-\textbf{x}|=\Delta x}{\beta_{\xi(\textbf{x})\xi(\textbf{z})}} -\sum_{\textbf{z}:|\textbf{z}-\textbf{y}|=\Delta x}{\beta_{\xi(\textbf{y})\xi(\textbf{z})}}\right\}
  \end{split}\text{.}
\end{equation}
The factor $\exp\{-8/3\Delta d\}$ just rescales the time scale of
migration $\tau$.

\begin{figure}[ht!]
 \centering
 \includegraphics[scale=0.40]{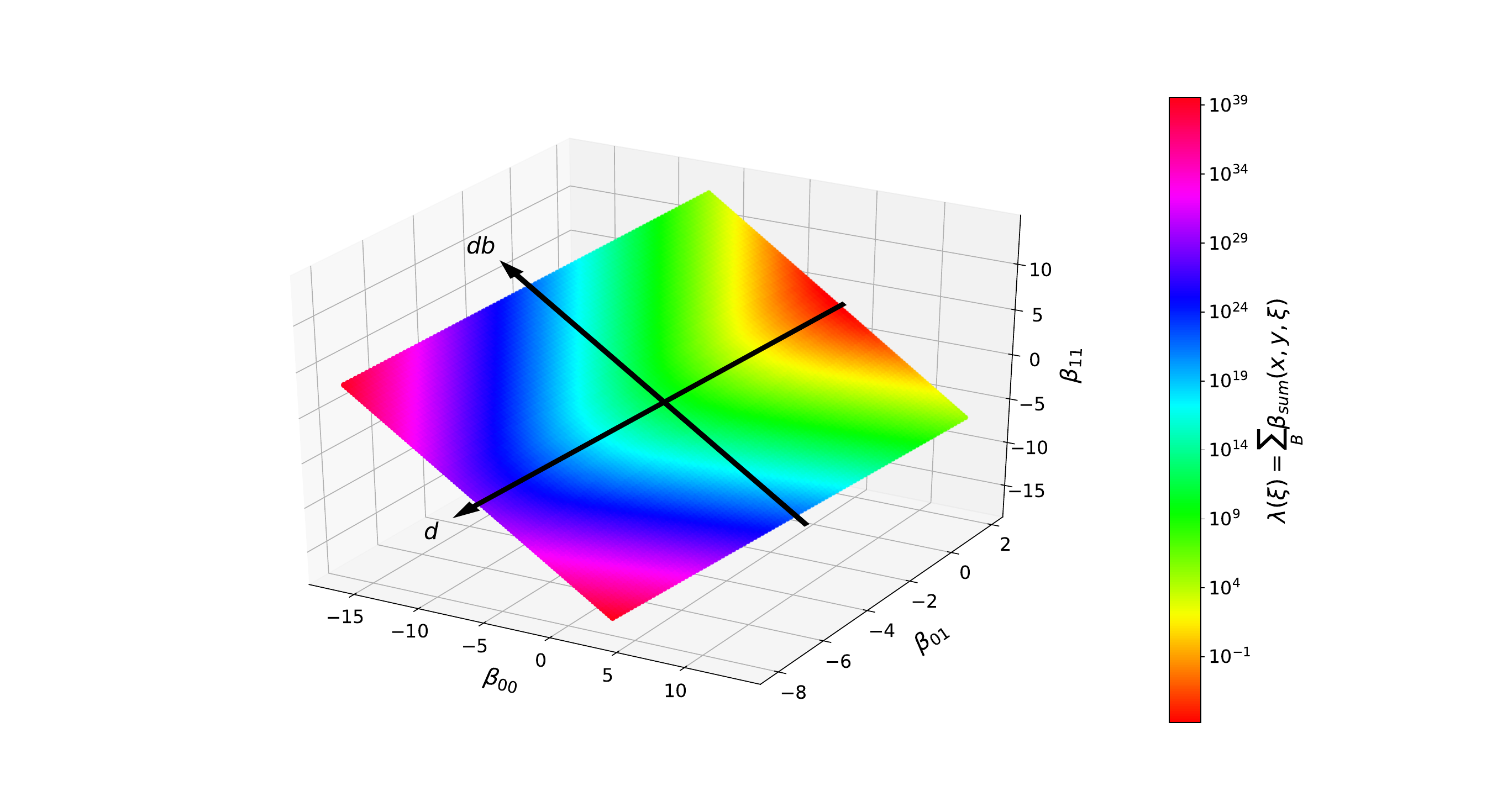}
 \caption{The influence on the time scale is trival for the parameter
   $d$, and non trival for the parameter $db$. Shown is, for a
   constant $\beta^*=3$ and random initial conditions $\xi$ with a
   $50/50$ cell type ratio, the color coded sum $\lambda_0$ of all
   heterotypic transitions rates. In direction of $(1,1,1)^T$, the
   value of $\lambda$ decreases and therefore the simulation time
   $\Delta t_\text{swap}$ for two neighboring heterotypic cells to
   change positions increases. In direction of $(-1,0,1)^T$, the time
   dependency is nontrivial, but symmetric to $db=0$.}
 \label{fig:main_Florian_paper_plot_4}
\end{figure}

The effects of the parameter $db$ on the model system are more complex
and have been examined numerically. We initialize with a random
configuration $\xi$ and measure successively, for each subsequent
configuration $\xi_t$ the sum $\lambda_t$ of all heterotypic
transition rates in the whole system at this time. The value
$\lambda_t$ sets the current average waiting time
$\Delta t_\text{swap} = 1/\lambda_t$ between two cell switches, see
implementation of the cellular automaton in SI text. We find that on
average an increase of the parameter $db$ will increase $\lambda_t$
and therefore decrease the average waiting time
$\Delta t_\text{swap}$. As illustration we show the dependency of
$\lambda_0$ on the parameters for a random configuration $\xi$ in
\figref{main_Florian_paper_plot_4}. Further, for a fixed parameter
$\beta^*$, an increase of $db$ will also increase the computing time, i.e.,
the number of cell switches required to reach the same level of
segregation~\cite{RosBoeLanVos2021}.

To further assess the influence of the initial configurations $\xi$, we
generated configuration with segregation indices $0.25$, as observed
initially in the experiments, by evolving a randomly mixed configuration
with different adhesion parameter sets up to this point and then
changed the adhesion parameters for further time. The comparison
between the segregation processes for the same adhesion parameters but
the different initial conditions, displayed in in
\figref{main_Florian_paper_plot_14_1} in SI, reveals a small influence
of the initial condition, but the scaling of the segregation indices is
not affected.

\subsection*{Segregation index}
As in the experiment, we use type specific segregation indices $\gamma_i$ to
determine the degree of segregation over time in the cell-based model.
For type $i \in W$, the index $\gamma_i$ is the average of the amount
$n_{\neq}(k)$ of heterotypic neighbors, where the average is taken over all
positions $k$ carrying cells of type $i$, in relation to the maximum possible
numbers of neighbors, which is $4$ for a von-Neumann neighborhood,
\begin{equation}
  \label{eq:gamma_definition}
  \gamma_i=\frac{1}{4}\left\langle
    n_{\neq}(k)\right\rangle_{\substack{k\in S \\
      \eta(k)=i}}=\frac{1}{4N_i}\sum_{\substack{k\in S \\
      \eta(k)=i}}{n_{\neq}(k)}=\frac{1}{4N_i}I \text{,}
\end{equation}
where $N_i$ denotes the total number of cells of type $i$ and $I$
denotes the interface length,
\begin{equation}
  \label{eq:interface_length_definition}
  I=\sum_{\substack{k\in S \\
      \eta(k)=i}}{n_{\neq}(k)}\text{,}
\end{equation}
which is another commonly used measure of segregation. Further, if an even
cell type ratio is given ($50/50$), it applies $N_i=N^2/2$, where
$N^2=\vert S\vert$. The resulting prefactor $2N^2$ is equal to the maximum
achievable interface length in the cellular automaton, which corresponds
to a checkerboard configuration where each cell has four heterotypic
neighbors, $I_\text{max}=2N^2$. Based on this, the relative interface
length $I_r$ can be defined as the interface length $I$ normalized by
the maximal interface length $I_\text{max}$,
\begin{equation}
  \label{eq:rel_interface_length_definition}
  I_r= \frac{I}{I_\text{max}}= \frac{1}{2N^2}\sum_{\substack{k\in S \\
  \eta(k)=i}}{n_{\neq}(k)}\text{.}
\end{equation}
Thus, for an even cell
type ratio $N_0=N_1$ the relative interface length $I_r$ is equal to
the segregation indices $\gamma_0=\gamma_1=I_r$. If the numbers of
cells of each type $N_i$ are not equal, it follows from \eqref{gamma_definition}
that the segregation indices $\gamma_i$ are inverse-proportional to
the cell type ratio
\begin{equation}
  \label{eq:gamma_celltyperatio}
  \frac{\gamma_0(t)}{\gamma_1(t)}=\frac{N_1}{N_0}\text{.}
\end{equation}
Therefore, the scaling exponents of $\gamma_i$ and $I_r$ are always identical.

Based on \eqref{gamma_definition} it is possible to calculate
the minimal segregation indices for a given field $N^2$. Since we only use a
quadratic field with periodic boundary condition for our simulations, the
minimal interface length can be assumed to be $I_\text{min} \approx 2N$.
For the corresponding segregation indices it follows:
\begin{equation}
  \label{eq:min_gamma}
  \gamma_\text{i,min}=\frac{1}{4N_i}I_\text{min}\approx\frac{N}{2N_i}\text{.}
\end{equation}
Note, that for equal cell ratio $N_i = \frac{N^2}{2}$ this lower boundary
scales inversely with the system size $\gamma_\text{i,min} \approx N^{-1}$.

To determine the goodness-of-fit for the segregation indices, we calculate the
averaged mean squared deviation with the following algorithm:

\begin{enumerate}
  \item Choose $50$ time points evenly on a logarithmic scale within the
  relevant time interval.
  \item Determine the corresponding values of the experimental segregation
  indices by piecewise linear interpolation between the discrete observation
  points of the experiment.
  \item For each cell type, HaCaT, PFK or EPC, average the two experimental time
  series evaluated at the above chosen 50 time points.
  \item Determine the squared deviation per point of the averaged experimental
  data and the corresponding value of the simulation.
  \item Average the squared deviation over both cell types for the each
  experiment.
\end{enumerate}

The previous equations \eqref{gamma_definition} and
\eqref{gamma_celltyperatio} apply exactly for periodic boundary conditions.
For other boundary conditions, the cell type ratio still
approximates the type specific segregation indices ratio
$\gamma_0(t)/\gamma_1(t)\sim N_1/N_0$, and the segregation indices
approximate the relative interface length $I_r \sim \gamma_i$. This is
due to the fact that boundary cells at the edge and in the corners have less
than $4$ neighbors, but their contribution gets less with rising
lattice size $N$, since the boundary size scales with $O(N)$ and the
lattice size scales with $ O(N^2)$.

\begin{figure}[ht!]
 \centering
 \includegraphics[scale=0.335]{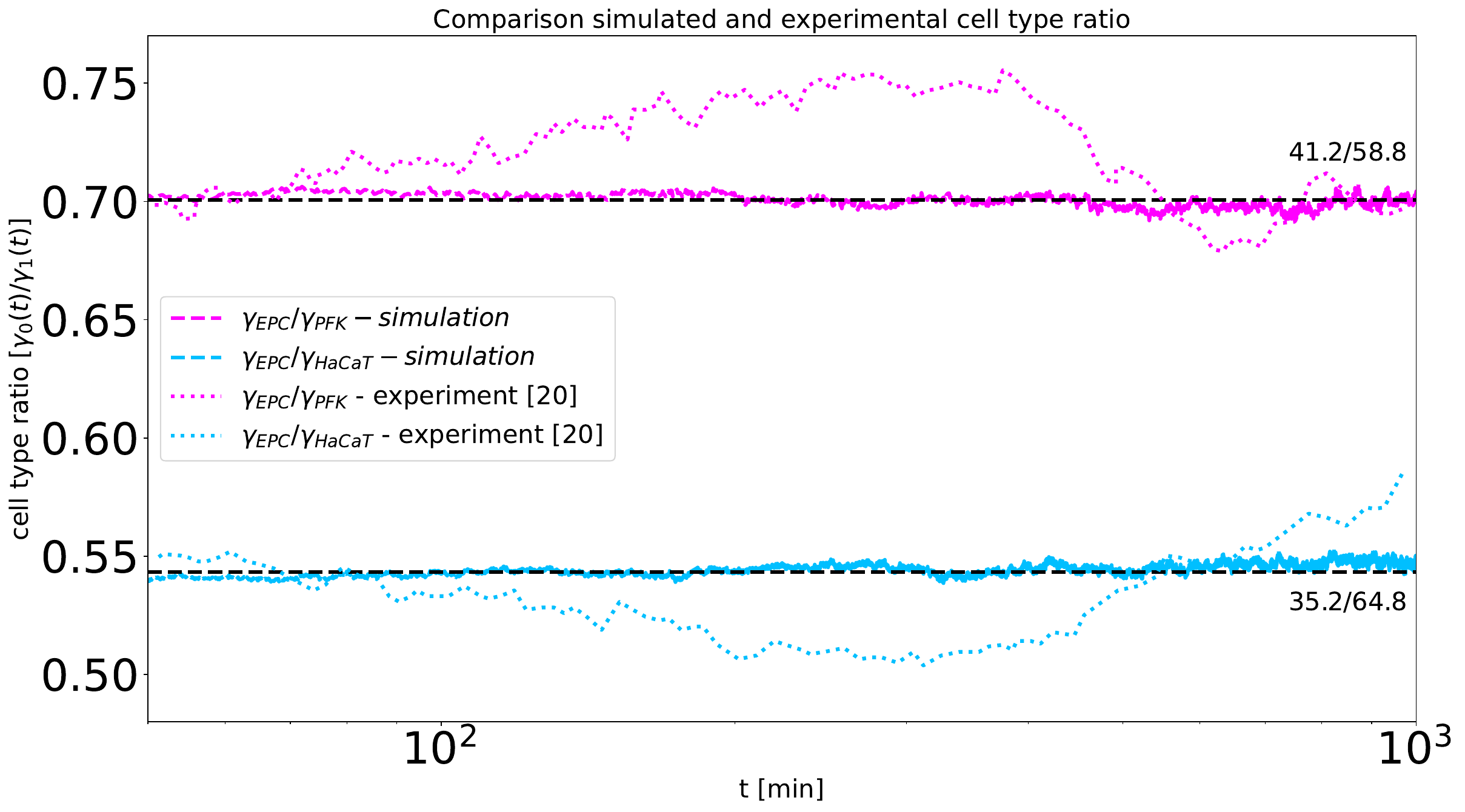}
 \caption{The cell type ratio for the simulation can be obtained from
   the experiments of Méhes et al.~\cite{MehMonNemVic2012}. Shown is a
   comparison of the cell type ratio $r=\gamma_0(t)/\gamma_1(t)$ from
   the experiments of Méhes et al.~\cite{MehMonNemVic2012} (dotted in
   color) and from the cellular automaton (dashed in corresponding
   color). The cell type ratio in the cellular automaton is set to the
   approximate mean of the ratios observed in the experiment (dashed
   black lines). The cell type ratio was calculated by the ratio of
   the type specific segregation indices for each time $t$ per
   experiment EPC with PFK (red) and HaCaT with EPC (blue).}
 \label{fig:main_Florian_paper_plot_7_1}
\end{figure}

Based on \eqref{gamma_celltyperatio} it is possible to calculate for
every pair of type specific segregation indices $\gamma_0(t)$ and
$\gamma_1(t)$ the corresponding cell type ratio and vice versa. We
define the cell type ratio as
\begin{equation}
  \label{eq:celltyperatio}
  r:=\frac{\gamma_0}{\gamma_1} \quad \text{with } N_0 \geq N_1 \text{,}
\end{equation}
where we assume without loss of generality that $r\leq1$.

In this sense, the cell type ratio in the experiment can be obtained
from the ratio of the corresponding segregation indices, see
\figref{main_Florian_paper_plot_7_1}. Indeed, the ratio of segregation
indices is relatively constant over the time span of the experiment,
and we set the cell type ratio of the cellular automaton accordingly,
as indicated in \figref{main_Florian_paper_plot_7_1}. In their
experiments, Méhes et al.~\cite{MehMonNemVic2012} choose the
cell type ratio such that initially the same amount of area is covered by
each type. Since the cells of each type are similar in size, as EPC is
$300\mu m^2$~\cite{FijEtal1983}, PFK is
$400 \mu m^2$~\cite{MehVic2013} and HaCaT is
$80-400\mu m^2$~\cite{BoeVePon1999}, it is reasonable to simulate the
segregation with the cellular automaton, where every cell has the same
space of the grid. However, the small differences in size imply that
the number of cells of each type in the experiment is not equal.
Instead, one estimates from the cell size ratios
$0.75=A_\text{EPC}/A_\text{PFK}=A_\text{EPC}/A_\text{HaCaT}$, cell
type ratios which are consistent with the ones obtained from the ratio
of segregation indices $0.70=N_\text{PFK}/N_\text{EPC}$ and
$0.54=N_\text{HaCaT}/N_\text{EPC}$ for EPC with PFK and HaCaT with
EPC, respectively, see \figref{main_Florian_paper_plot_7_1}. Note that
the specific cell type sizes where not reported by Méhes et
al.~\cite{MehMonNemVic2012} and cells can vary in size during an
experiment as well as depending on the experimental setup.

Additionally to the segregation indices and the interface length, the
average cluster diameter is a third commonly used measure to determine
order in segregation processes. For the cellular automaton it can be
shown that the average cluster diameter $d$ is inverse-proportional to
the interface length $I$, assuming the cell type ratio equals $50/50$,
the cluster size distribution is narrow, i.e.
$\langle d_l \rangle^2 \approx \langle d_l^2 \rangle$, the total area
$A_\text{sum}$ of all clusters is constant, and the clusters are
approximately circular. In the following $n_c$ denotes the number of
clusters, $A_l, l\in\mathbb{N}|1\leq l\leq n_c$ the size and $U_l$ the
scope of the $l$-th cluster. Approximating the clusters as circles, we
have
\begin{equation}
  \label{eq:cluster_1}
  \begin{split}
    & \sum_{l}^{n_c}{A_l} = A_\text{sum} = n_c \langle A_l \rangle = n_c \frac{4}{\pi}\langle d_l^2 \rangle \Leftrightarrow n_c=\frac{4A_\text{sum}}{\pi\langle d_l^2 \rangle}\\
    & \sum_{l}^{n_c}{U_l} = I = n_c\langle U_l \rangle = n_c\pi\langle
    d_l \rangle \Leftrightarrow n_c = \frac{I}{\langle d_l \rangle\pi}
    \text{.}
  \end{split}
\end{equation}
By combining the two expressions for $n_c$ in \eqref{cluster_1}, we get
\begin{equation}
  \label{eq:cluster_2}
  \langle d_l \rangle = \frac{I\langle d_l^2 \rangle}{4A_\text{sum}}
  \Rightarrow \langle d_l \rangle \underset{\langle d_l \rangle^2
    \approx \langle d_l^2 \rangle}{\approx} \frac{\langle d_l^2
    \rangle}{\langle d_l \rangle} = \frac{4 A_\text{sum}}{I}\text{,}
\end{equation}
where the last approximation is only valid for a narrow distribution
of cluster sizes. Since $4A_\text{sum}$ is constant, it results that
the average cluster diameter is inverse-proportional to the interface
length $\langle d_l \rangle\sim 1/I$. We infer from the fact that the
average cluster diameter in both, the cellular automaton and the
Cahn-Hilliard model, is inverse-proportional to the interface length,
that their distribution of cluster sizes is sufficiently narrow. This
is consistent with recent observations for the CPM~\cite{Dur2021},
where the same inverse-proportional behavior is observed
asymptotically when the formed clusters are approximately circular.

\section*{Conclusion}

By calibrating a 2D cellular automaton model which solely
incorporates differential adhesion to the experimental setup of Méhes
et al.~\cite{MehMonNemVic2012}, we reproduce experimentally observed
segregation indices. While Méhes et al. interpreted the decay of the
experimental segregation indices as an algebraic scaling with the
exponent of $1/3$, the cellular automaton model exhibits a logarithmic
decay at the time scale of the experiment, as it belongs to the
transitory regime of the model. Since Steinberg~\cite{Ste1970} also
proposed that cell segregation is similar to that of fluids, we
additionally compare the experimental results with fluid segregation
expressed by the 2D Cahn-Hilliard model. By developing a mapping
between the cellular automaton model and the Cahn-Hilliard model, only
one parameter remains to be fitted. The resulting segregation indices
from the Cahn-Hilliard model fit the experimental ones well, although
they rather follow an pseudo-algebraic decay with exponent $1/4$ than $1/3$
on the relevant time interval. The match of the experimental results
with both models highlights the possible ambiguity of scalings on
the short time spans
of the experimental data. Our results also emphasize that the transitory
regime of these models is relevant for the spatio-temporal scales of the
experiment rather than the asymptotic regime. This is in contrast to most
of the theoretical studies, which usually focus on the asymptotic regime
of the cell segregation models and do not calibrate the model parameters
to the physical constraints of the experiments.

Our results highlight the importance of additional metrics to compare
segregation between simulations and experiments, in order to avoid the
ambiguity of scaling laws on the limited time spans of the experiments.
Thus, future experiments on cell segregation should report their
observations in terms of several metrics, like segregation indices,
cluster size distribution and average cluster diameter, and provide
the raw data to allow further retroactive analysis in comparison
with simulations.

While our focus here is segregation in 2D experiments and models, it
would be interesting to extent our approach to 3D tissues.
Cochet-Escartin et al.~\cite{CocLocSteCol2017} studied segregation in
3D tissue over half an order of magnitude of time. They measured an
algebraic decay with exponent $0.74$ for the segregation indices in
the experiment and $1/2$ for that in a corresponding CPM model. Note that the
measured algebraic decay is only displayed for a quarter order of
magnitude in time. However, this is remarkable, since the exponent of
the algebraic decay in a 3D space should rather decrease, compared to
a 2D space according to the diffusion-coalescence
mechanism~\cite{BeaAlmBru2017, Mea1990, Kol1984}. This discrepancy
suggests that the segregation was observed in the transitory regime,
which points to the importance of studying transitory regimes in 3D
tissues as well.

\section*{Acknowledgments}
We thank Tam\'{a}s Vicsek for his helpful communication about his
publication. The authors A.V.-B., S.L., H.-J.B., F.F. acknowledge that
this research is co-financed by the EU, the European Social Fund (ESF) and
by tax funds on the basis of the budget passed by the Saxon state parliament.
Further, A.V.-B and F.F. acknowledge support by Sächsisches
Staatsministerium für Wissenschaft und Kunst (SMWK) project FORZUG II TP 3.
The funders had no role in study design, data collection and analysis, decision
to publish, or preparation of the manuscript.

\newpage

\renewcommand{\theequation}{S\arabic{equation}}
\setcounter{section}{1} 
\setcounter{subsection}{0} 
\renewcommand{\thesubsection}{\Alph{subsection}}
\setcounter{equation}{0}  
\renewcommand{\thefigure}{\Alph{figure}}
\setcounter{figure}{0}  
\renewcommand{\thetable}{\Alph{table}}
\setcounter{table}{0}  

\section*{Supporting Information (SI)}

\begin{figure}[ht!]
 \centering
 \includegraphics[scale=0.3]{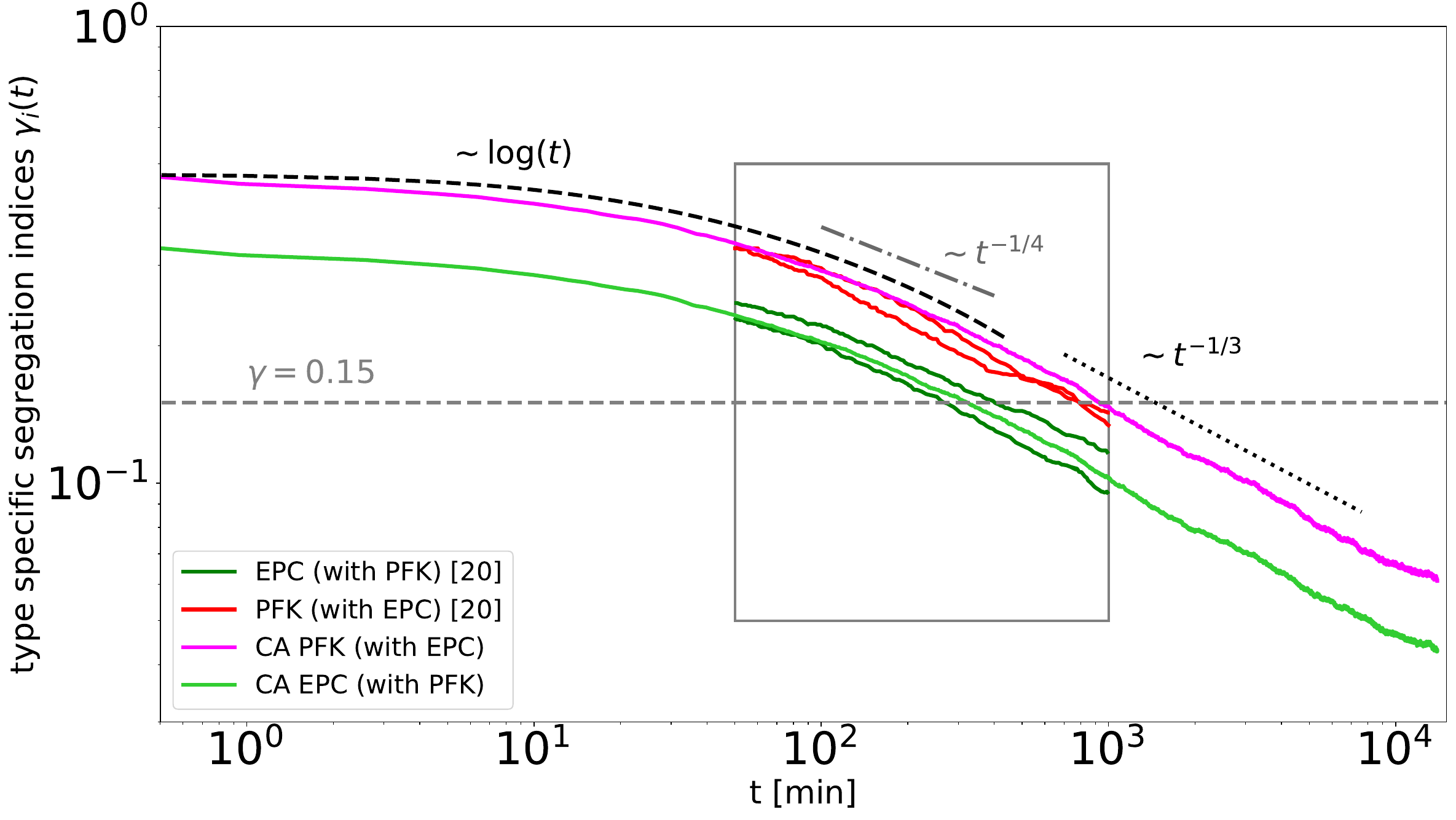}
 \caption{The data of Méhes et al.~\cite{MehMonNemVic2012} fall into
   the transitory regime of the cellular automaton. This is shown
   exemplary for the segregation of PFK with EPC. The simulation with
   the same parameters as in \figref{main_Florian_paper_plot_1_4_1},
   with time scale of migration $\tau_\text{PFK-EPC}\approx 4.2$ min,
   cell type ratio $N_\text{PFK}/N_\text{EPC}=41.2/58.8$, adhesion
   parameters
   $(\beta_\text{PFK-PFK},\beta_\text{EPC-PFK},\beta_\text{EPC-EPC})=(-8.06,-6.56,-0.06)$,
   and $140^2$ cells is run for an additional order of magnitude in
   time such that the segregation indices drop down to
   $\gamma\approx0.05$, which is much smaller than the smallest
   indices $\gamma\approx0.1$ in
   \figref{main_Florian_paper_plot_1_4_1}. Within the given time
   interval of the experiments (gray box), the cellular automaton
   shows ambiguous scaling behaviors. The segregation decays
   logarithmically at the beginning ($\gamma_i \sim 0.58 \log(t)$,
   black dashed line), followed by an algebraic decay with exponent of
   $1/4$ (gray dash-dotted line) in the transitory regime, and finally
   by a algebraic decay with exponent $1/3$ (black dotted line) at
   smaller segregation indices $\gamma_{\text{EPC}}<0.15$, which could
   be considered as the asymptotic regime.}
 \label{fig:main_Florian_paper_plot_10_2}
\end{figure}

\begin{figure}[ht!]
 \centering
 \includegraphics[scale=0.28]{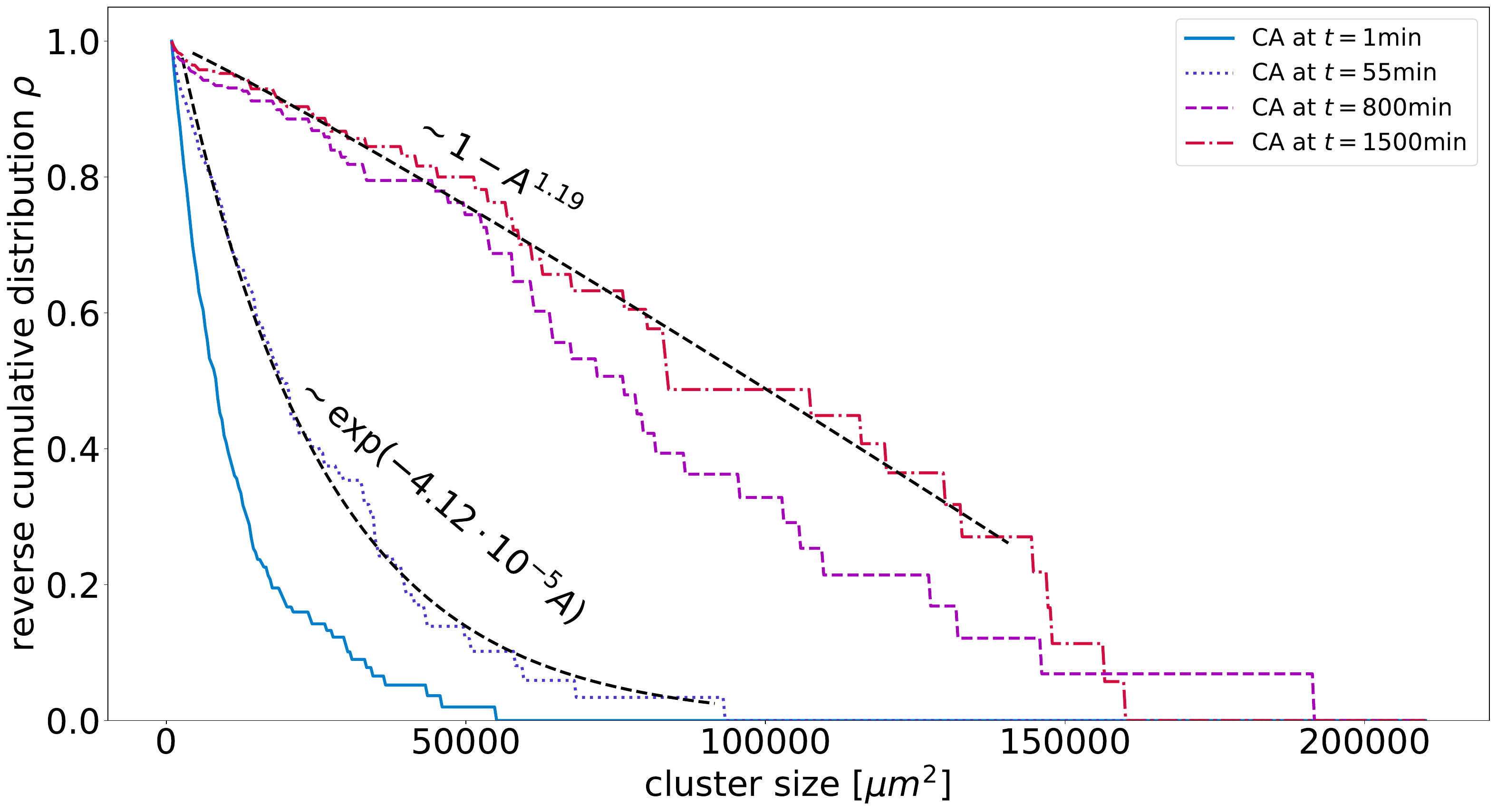}
 \caption{The cluster size distribution $\rho$ in the cellular automaton
   displays different characteristics for the early and the
   later regime. The reverse cumulative cluster size
   distributions are shown at four points of segregation. Note that for
   the early regime, $t=1$min and $t=55$min, the reverse cumulative cluster
   size distribution $\rho$ follows an exponential decay $\exp(-k\cdot A)$. For
   the later regime, $t=800$min and $t=1500$min, the cluster size
   distribution displays an algebraic decay with exponent $\approx 1$.}
 \label{fig:main_Florian_paper_plot_11_2}
\end{figure}

\subsection*{Cahn-Hilliard Navier-Stokes}

The Cahn-Hilliard Navier-Stokes model accurately describes the evolution of
two immiscible fluids under flow and
diffusion~\cite{AlaBodHahKliWeiWel2013, AlaVoi2012}. The model is also used
as a typical choice for simulating fluid
segregation~\cite{WitBacVoi2012, VooGli1988, Voo1985, RogDes1989}. This
model is known for producing an algebraic scaling with the exponent of $1/3$
for segregation processes with small length
scale~\cite{GarNieRum2003, NasNar2017}. For larger length scales, an
exponent of 2/3 can be observed~\cite{NasNar2017}. The complete model is
given by the following set of differential equations:
\begin{equation}
  \label{eq:CHNS}
  \begin{aligned}
    & \partial_t\Phi+\textbf{u}\cdot\nabla\Phi = D\Delta\mu \text{,}\\
    & \mu = -\epsilon^2\Delta\Phi+\Phi^3-\Phi \text{,}\\
    & \rho(\partial_t\textbf{u} + (\textbf{u}\cdot\nabla)\textbf{u}) + \nabla p = \nabla \cdot(\eta(\nabla \textbf{u}+\nabla \textbf{u}^T)) + \tilde{\sigma}\epsilon^{-1}\mu\nabla\Phi \text{,}\\
    & \nabla \cdot \textbf{u} = 0\text{.}
  \end{aligned}
\end{equation}
At small length scales, predefined by the size of the biological cells, the
fourth order diffusion in the Cahn-Hilliard equations dominates, such that
the influence of flow can be neglected and \eqref{CHNS} simplifies to:
\begin{equation}
  \begin{aligned}
    &\partial_t\Phi = D\Delta\mu \text{,}\\
    &\mu = -\epsilon^2\Delta\Phi+\Phi^3-\Phi \text{.}
  \end{aligned}
\end{equation}
These equations are defined for a domain $\Omega=[0,L_x]\times[0,L_y]$ where
$L_x$ and $L_y$ denotes the maximal size of the domain. We define a phase
$\Phi:\Omega\rightarrow[-1,1]$ on this domain, where $\Phi\approx 1$ denotes
the first fluid, like water, and $\Phi\approx -1$ denotes the second fluid,
like oil. Values of $|\Phi| < \Phi_0$ are defined as interface area, e.g.
$\Phi_0=0.9$. The width of the interface area is proportional to the
parameter $\epsilon$:
\begin{equation}
    \label{eq:main_Florian_paper_plot_2}
    \delta = \text{arctanh}(\Phi_0)\sqrt{2}\epsilon\text{,}
\end{equation}
see \figref{main_Florian_paper_plot_2} below.

\begin{figure}[ht!]
 \centering
 \includegraphics[scale=0.3]{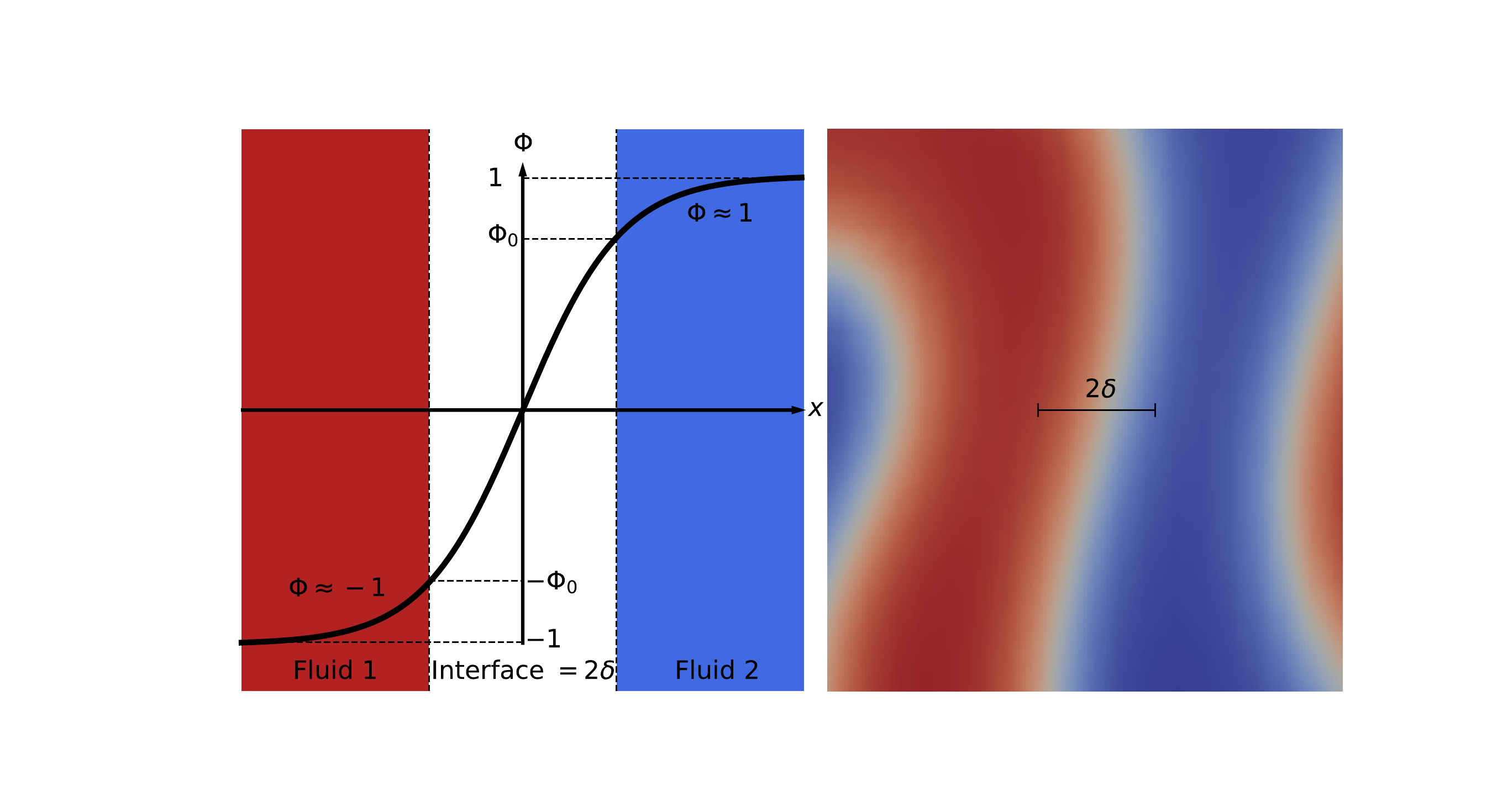}
 \caption{Schematic representation of the relationship between $\Phi$ and
 $\delta$ and their correlation to the interface area.}
 \label{fig:main_Florian_paper_plot_2}
\end{figure}

The parameter $D$ is the mobility constant and influences the time
scale of the diffusion process, which is set by $\tau \sim \frac{\epsilon^2}{D}$.
With the phase $\Phi$ and the
parameter $\epsilon$ the chemical potential $\mu(\Phi, \epsilon)$ can
be calculated. Each fluid has its own typical parameters like
viscosity $\eta_{\Phi=-1}$ and $\eta_{\Phi=_1}$ and density
$\rho_{\Phi=-1}$ and $\rho_{\Phi=_1}$. By linear interpolation of the
viscosity and density pairs, the functions $\eta(\Phi)$ and
$\rho(\Phi)$ can be calculated. Dependent on the types of fluids there
is a surface tension $\sigma$, which enters the Navier-Stokes equations
after a rescaling as
$\tilde{\sigma}:=\frac{3}{2{\sqrt{2}}}$~\cite{AlaBodHahKliWeiWel2013, AlaVoi2012}.

The implementation of this model follows a special pressure projection scheme
with incomplete pressure iterations and an explicit Euler approach described
in the paper of Adam et al.~\cite{AdaFraAla2020}.

The interface length $I$ for this model is approximated by the Cahn-Hilliard
surface energy,
\begin{equation}
    \label{eq:interface_ch}
    I\approx\frac{3}{2\sqrt{2}}\int_{\Omega}{\frac{1}{\epsilon}W(\Phi)+\frac{\epsilon}{2}|\nabla\Phi^2|}\text{,}
\end{equation}
\begin{equation}
    W(\Phi)=\frac{1}{4}(1-\Phi^2)^2\text{.}
\end{equation}

At the start of a simulation, the phase $\Phi$ is set randomly in between $0$ and $1$
following a uniform distribution for each grid point. Then an initial settling
process takes place, where no clear phases are observed, since all  values of
$|\Phi|$ are significantly smaller than $1$. In this time span, the measured interface
length with \eqref{interface_ch} is irrelevant for our purpose, since there are no clear
phases which segregate. This effect is displayed in \figref{main_Florian_paper_plot_9}
A)-C). If \eqref{interface_ch} yields a plateau over some time, the settling
process ceases, two well-mixed phases have established, and the CH behavior can
be interpreted as segregation process. Note, that one can also observe a slight
local increase of the quantity $I$ in \eqref{interface_ch} instead of a plateau.

\subsection*{Mapping of the cellular automaton model and the Cahn-Hilliard model}

\begin{figure}[ht!]
 \centering
 \includegraphics[scale=0.2]{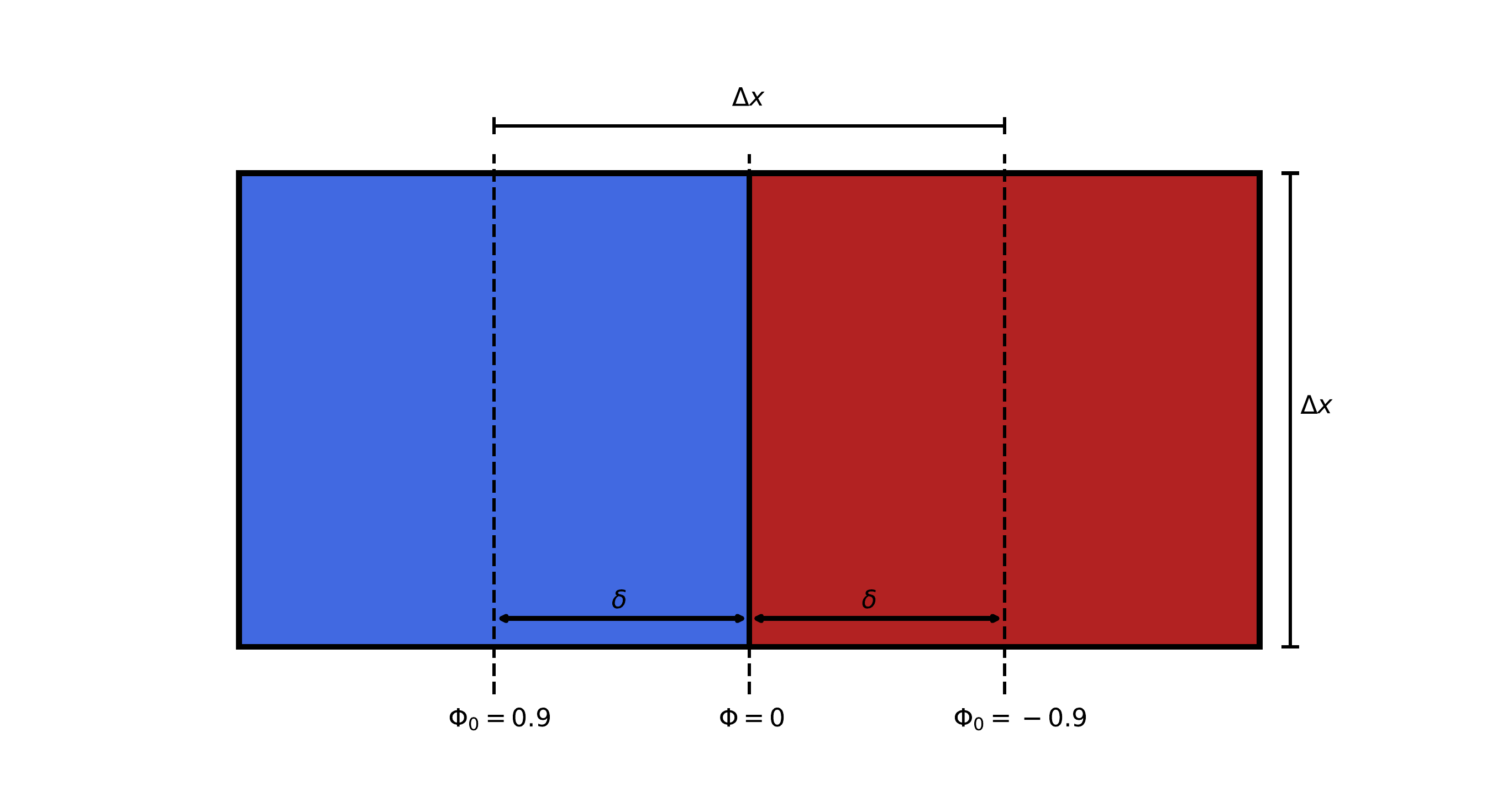}
 \caption{Visualization of the mapping of the transition area between
 the cellular automaton model and the Cahn-Hilliard model. Each square
 represents a cell in the cellular automaton, where red denotes $i=0$
 and blue $i=1$ for $i\in W$. The length of one side is defined as
 $\triangle x$ and can be calculated by the square root of the specific
 cell type area $A_0,A_1$. We define the middle of each cell to be
 equally to each phase $\Phi_0 = 0.9,-0.9$ of the Cahn-Hilliard model.
 Therefore can the absolute width of the transition area $\delta$ be
 calculated and $\epsilon$ can be determined.}
 \label{fig:main_Florian_paper_plot_8}
\end{figure}

We developed a mapping process, to equally start a simulation of the cellular
automaton model and the Cahn-Hilliard model and to compare their segregation
behaviors. Therefore  a match of the time or length scale between both models
is needed. Since the cellular automaton initially always segregates
logarithmically and the Cahn-Hilliard model algebraically, the time scale can
never be the same. However, the length scale can be matched. The length scale
in the Cahn-Hilliard model is set by the parameter for the width of the
interface area $\epsilon$. Since the interface area in the cellular automaton
is sharp, we define an equally wide transition to be from the middle of one
cell $\Phi\approx -0.9$ to the middle of a neighboring heterotypic cell
$\Phi\approx 0.9$. The length of a side $\triangle x$ of one cell in the
cellular automaton can be calculated by the square root of the average of
the specific cell areas $A=(A_0+A_1)/2\approx 350\mu \text{m}^2$. The transition
area in the cellular automaton refers to $2\delta$, see
\eqref{main_Florian_paper_plot_2}, and is equal to one cell side length
$\triangle x\approx\sqrt{350}\mu \text{m}$, see \figref{main_Florian_paper_plot_8}.

\begin{figure}[ht!]
 \centering
 \includegraphics[scale=0.45]{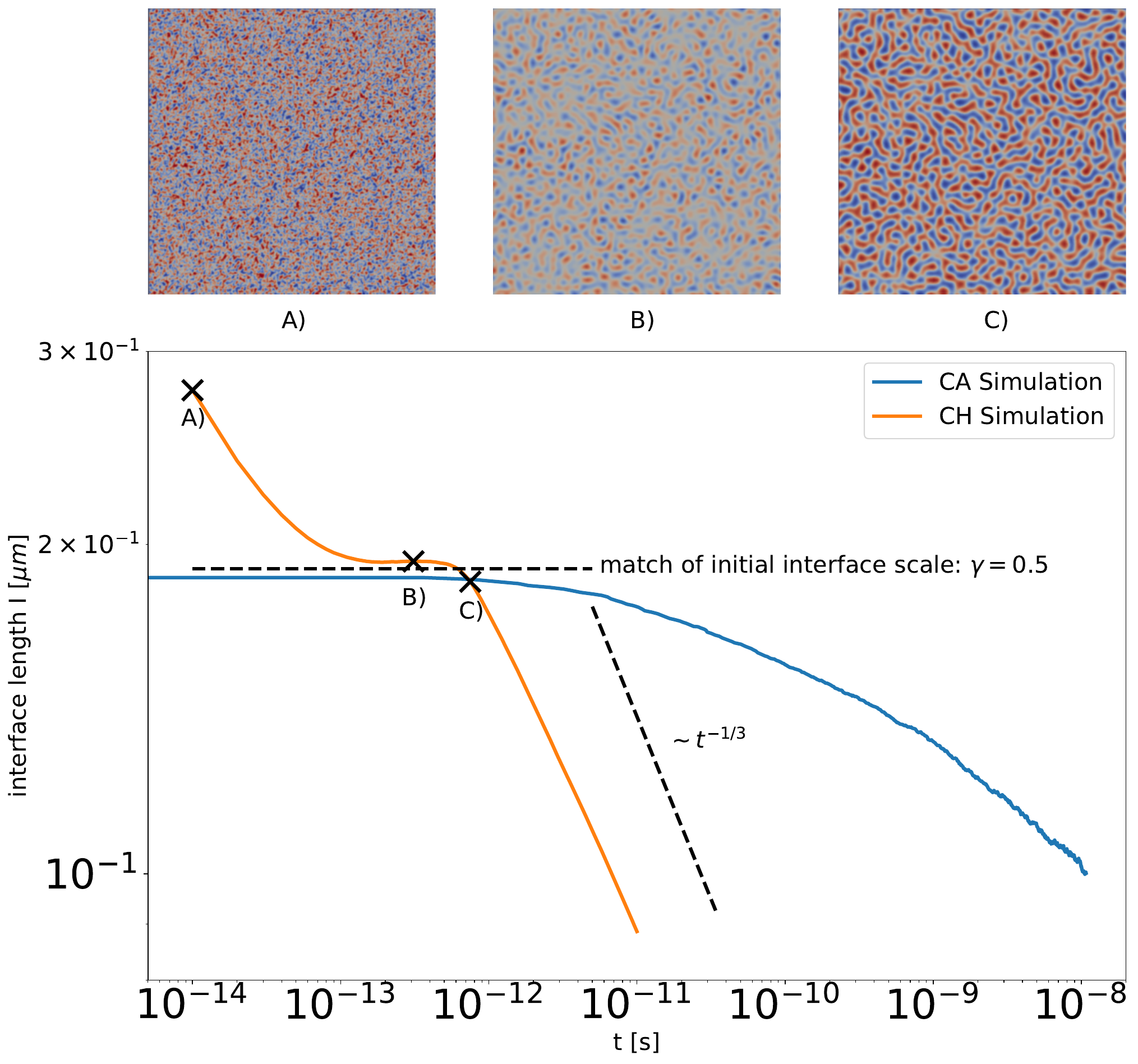}
 \caption{The cellular automaton and the Cahn-Hilliard model after an initial
 settling process start at the same interface length. The simulation
 of the cellular automaton is started with a random cell mixture, adhesion
 parameter $\textbf{$\beta$}=(-1.56, -3.06, -1.56)^T$, time scale of
 migration $\tau=1\text{s}$, $100^2$ cells, and a cell type ratio of $50/50$.
 The Cahn-Hilliard simulation is started from a random, uniformly distributed field
 with a mobility constant $D=0.24\mu\text{m}^2/\text{s}$. Note that the
 Cahn-Hilliard model has always an initial settling process, where the
 phases get formed, see Fig A) and B), concluded by a plateau phase for the interface
 length where two well-mixed phases can be distinguish, see Fig C). Therefore,
 the interface length at point C) of the Cahn-Hilliards
 model is matched with the initial interface length in the cellular automaton.}
 \label{fig:main_Florian_paper_plot_9}
\end{figure}

Further, the cell type ratio of the cellular automaton can be directly
integrated into the Cahn-Hilliard model, by initializing the simulation
with an equal phase ratio. The domain size $L_x=L_y$ can be obtained by
the number of cells per dimension multiplied with the average size of a
cell $N\delta =N\sqrt{350}\mu \text{m}$. Only the mobility constant $D$ needs
to be fitted. Since the interface width $\epsilon$ is set, the time
scales are only influenced by $D$. Therefore, if $D$ is doubled, the
time scale is halved. If the length scales are matched, the initial interface
length $I$ will be equal in both models, if both start from a random field.
The comparison of both models can be seen in \figref{main_Florian_paper_plot_9}.

\subsection*{Two point correlation method}

\begin{figure}[ht!]
 \centering
 \includegraphics[scale=0.25]{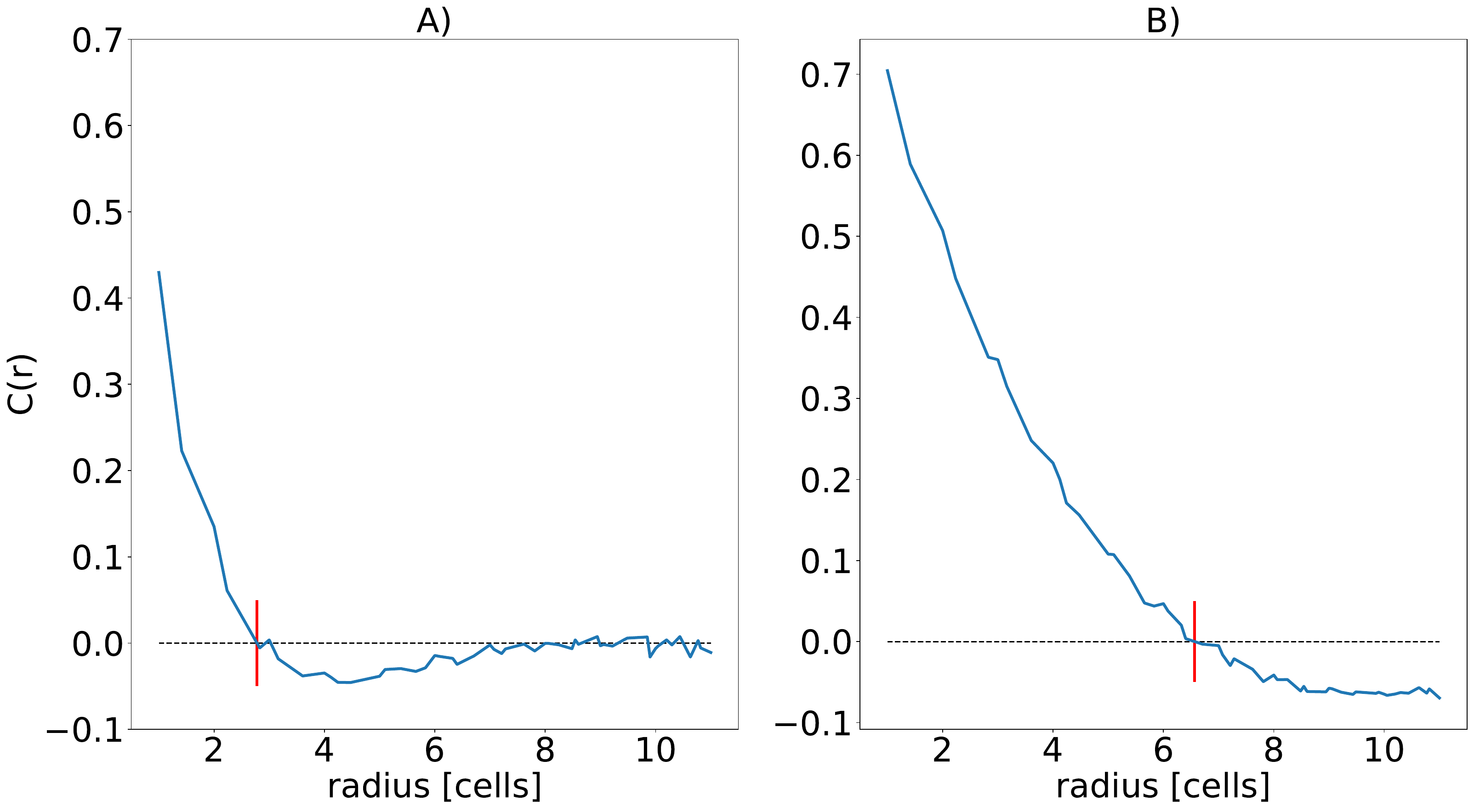}
 \caption{Exemplary development of the two point correlation in the cellular
  automaton. Panel A) shows the correlation $C(\textbf{r})$ at the time $t=55$min, which
  is similar to the start of the experiment. Panel B) shows the correlation
  $C(\textbf{r})$ at the time $t=800$min, which is similar to the end of the experiment.
  The radius $r$ is given in cellular automaton cells of width
  $\Delta x=\sqrt{350}\mu \text{m}$. The average radius of a cluster is defined
  by the smallest radius at which the correlation becomes zero $C(\textbf{r})=0$
  (red line).}
 \label{fig:main_Florian_paper_plot_12_1}
\end{figure}

The two point correlation method is commonly used to measure the
average cluster diameter of cells~\cite{MehMonNemVic2012, NakIsh2011}.
Our implementation follows

\begin{equation}
  \label{eq:2pointcorr}
  \begin{split}
    C(r,t) &= \int_0^{2\pi}C(\textbf{r},t)d\theta \\
    C(\textbf{r},t) &= \langle \Phi(\textbf{r}_\textbf{0},t) \Phi(\textbf{r}_\textbf{0} + \textbf{r},t)
    \rangle_{\textbf{r}_\textbf{0}} - \langle
    \Phi(\textbf{r}_\textbf{0},t) \rangle^2_{\textbf{r}_\textbf{0}}
  \end{split}
\end{equation}

where the phase $\Phi(\textbf{r},t)$ is 1 if the first cell type is
present at $\textbf{r}$ and $-1$ if the other is present,
and $\langle\text{.} \rangle_{\textbf{r}_\textbf{0}}$ denotes the
average over all grid points $\textbf{r}_\textbf{0}$. The average
radius of a cluster is defined by the smallest radius at which the
correlation becomes zero $C(\textbf{r},t)=0$. Two examples of the
correlation function can be seen in \figref{main_Florian_paper_plot_12_1}.

\subsection*{Video analysis}

\begin{figure}[ht!]
 \centering
 \includegraphics[scale=0.3]{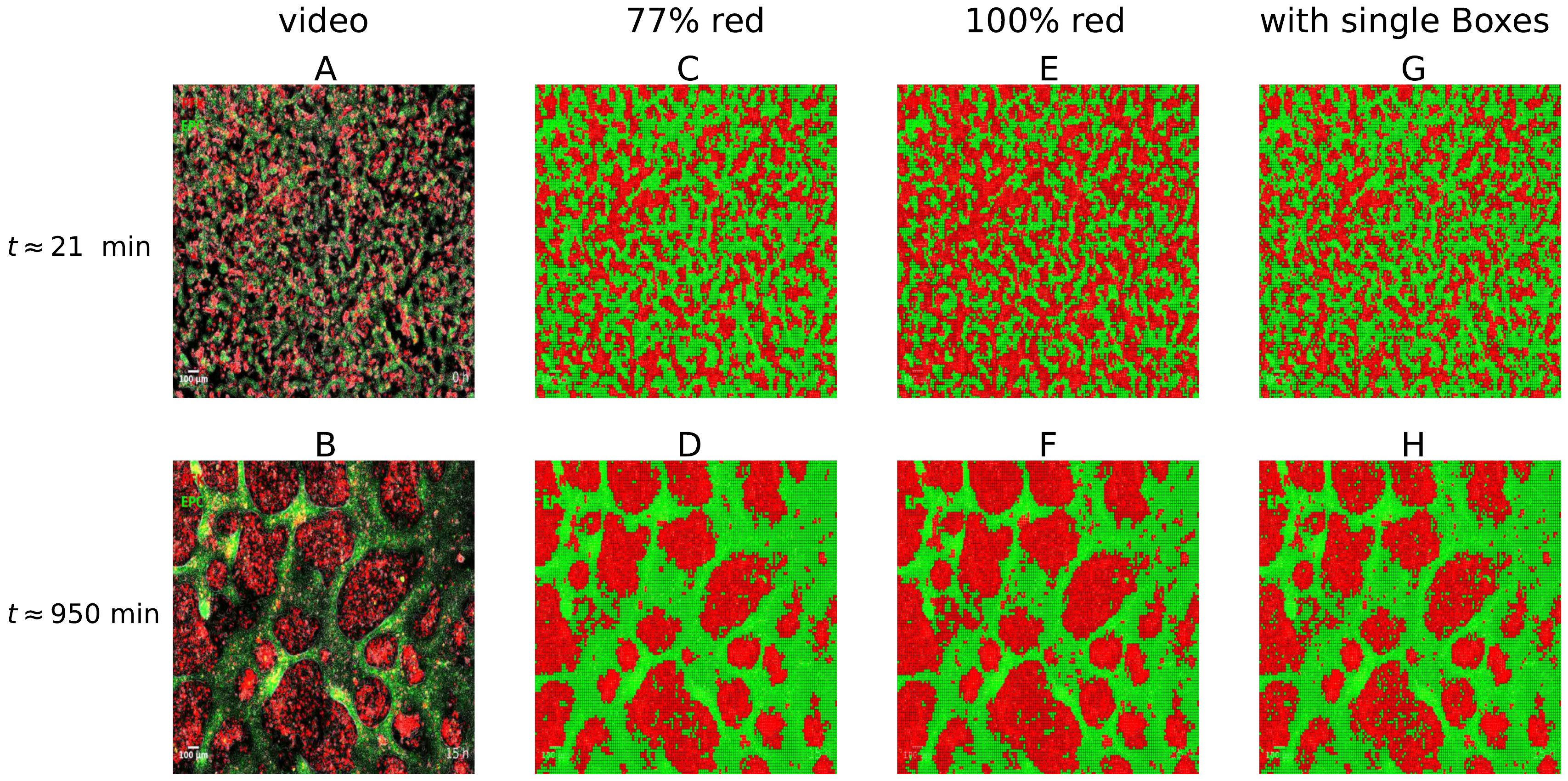}
 \caption{Illustration of the color calibration method used in the video
   analysis. The first row shows a frame from the experiment PFK with
   EPC at the time $t = 21$ min. The second row shows a frame at the
   time $t=950$ min, of the same experiment. A) and B) display the
   original frame of the video S5 of Méhes et
   al.~\cite{MehMonNemVic2012}. C) and D) show the result of the video
   analysis with the factor $77\%$ for the red channel and when single
   boxes in opposite-type environment are removed. E) and F) show the
   result when the color values per box are treated equally, which
   corresponds to a factor of $100\%$ for the red channel. G) and H)
   show the effect when individual boxes in opposite-type environments
   are not removed but a factor of $77\%$ for the red channel is used.
   Comparison of the second and third column justify that the red (PFK)
   pixel color sum value for each box should be multiplied by $77\%$
   and than compared to the green (EPC) pixel color sum value.
   Comparison of the second and fourth column with the first one
   justify the removal of single boxes in opposite-type environment.}
 \label{fig:main_Florian_paper_plot_11_5}
\end{figure}

\begin{figure}[ht!]
 \centering
 \includegraphics[scale=0.25]{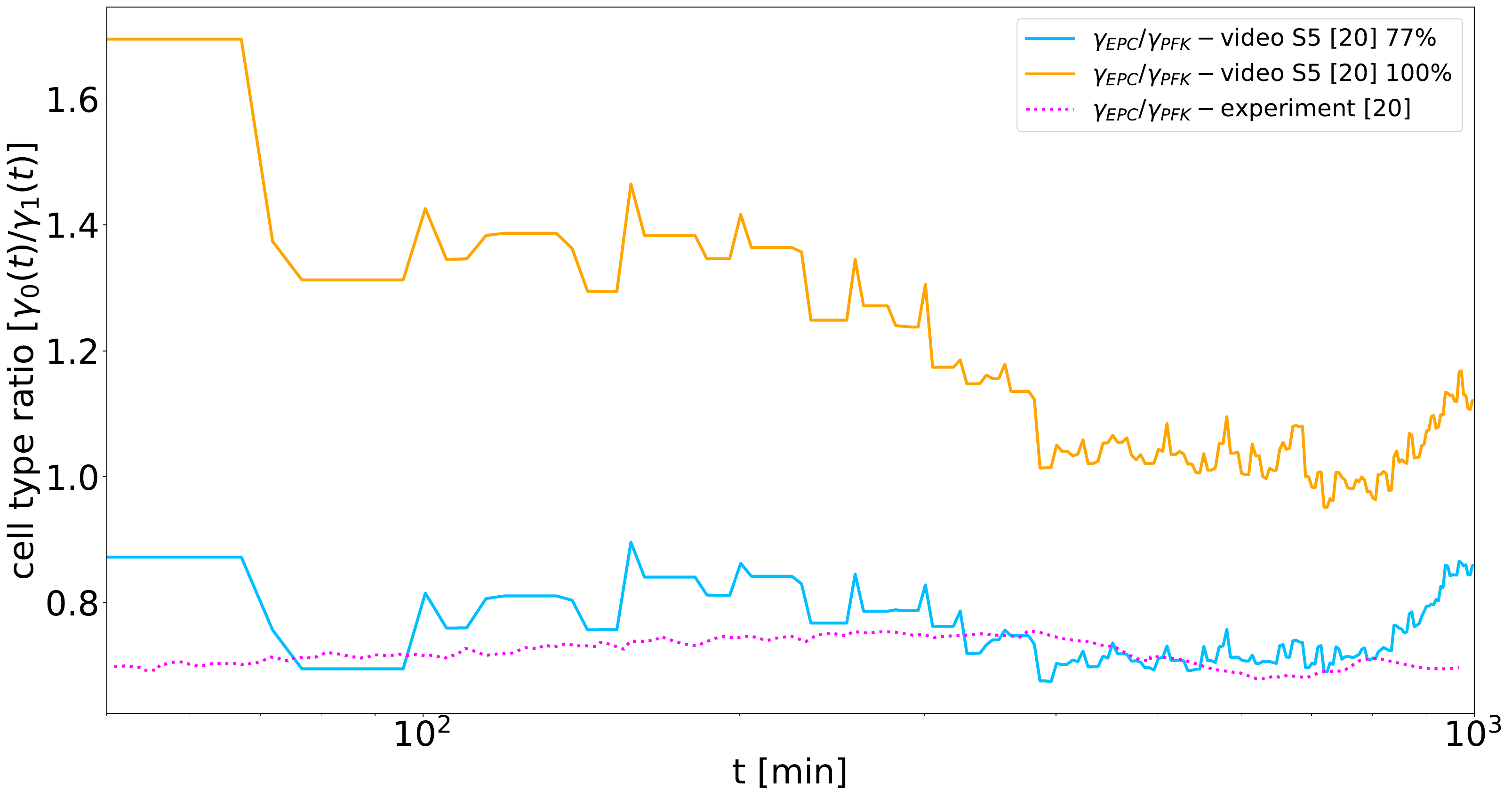}
 \caption{Illustration of the color calibration method used in the video
   analysis. The dotted magenta line shows the cell type ratio
   reported by Méhes et al.~\cite{MehMonNemVic2012}. The solid orange
   line represents the cell type ratio, if the pixel color sum for red
   (PFK) and green (EPC) is treated equally. The solid blue line
   displays the cell type ratio if the red (PFK) pixel color sum value for
   each box is multiplied by $77\%$ and than compared to the green
   (EPC) pixel color sum value. If both colors are treated equally
   (orange line), the cell type ratio varies over time, contradicting the fact
   that cell numbers in the experiment were kept constant. This supports
   the use of the $77\%$ factor for the red channel.}
 \label{fig:main_Florian_paper_plot_11_4}
\end{figure}

Since the details of the video analysis of Méhes et
al.~\cite{MehMonNemVic2012} are not fully available, we reanalysed the
video S5 documenting the segregation of PFK and EPC, see
\figref{main_Florian_paper_plot_1_4_3} center panels. Through the
depicted scale, we identified that $100\mu m$ corresponds to 37 pixel.
Since the average size of PFK and EPC cells equals $\sim 350\mu m^2$,
a square with a edge length of $\sqrt{350} \mu m\approx 7$ pixel
yields the same area. Further, we divided each frame in boxes of $7$ by
$7$ pixels and assigned a cell type to each box. To assign the cell type to each
box, we sum the red (PFK) and green (EPC) color channel of each pixels
RGB value for each box. We found, that favoring EPC cells in the
interpretation of the images by multiplying the red (PFK) sum value
with $77\%$ and classifying the entire box as red cell type (PFK) only if this
reduced value is greater than the non-reduced green (EPC) sum value leads to
stable cell type ratios over time, see \figref{main_Florian_paper_plot_11_4}.
In contrast, if both pixel color values were treated equally, then too many boxes
would be assigned to red (PFK) as \figref{main_Florian_paper_plot_11_5}
and \figref{main_Florian_paper_plot_11_4} highlight.

\begin{figure}[ht!]
 \centering
 \includegraphics[scale=0.25]{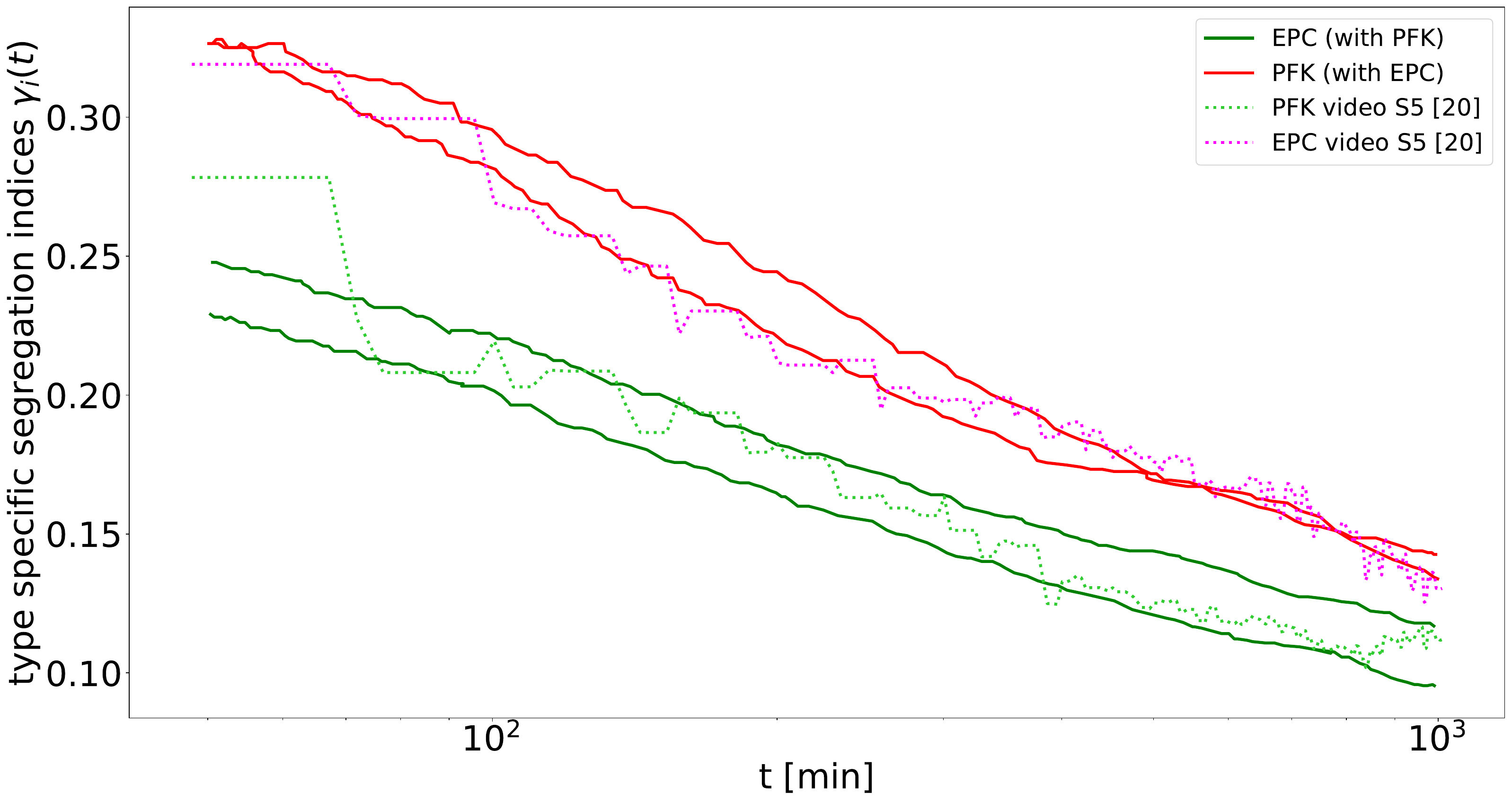}
 \caption{The segregation indices computed from our video analysis
   (dotted lines) fits well the indices reported by Méhes et
   al.~\cite{MehMonNemVic2012} (solid lines) for EPC (green) with PFK
   (red). The averaged mean squared deviation $\Delta\gamma$ of the result of
   the video analysis equals $0.0032$.}
 \label{fig:main_Florian_paper_plot_11_6}
\end{figure}

We further find, in addition to the $77\%$ rule, that single boxes,
which are surrounded by the opposite type, should be removed, to
gather a representative grid in regards to the original video. Without
the removal of single boxes, we find a biological incorrect grid
representation of the original experiment, as, especially in the later
stages, single boxes of EPC (green) can be seen in the larger clusters
of PFK (red), which is not the case for the video, see
\figref{main_Florian_paper_plot_11_4} B) and H). It follows, that with
the removal of single $7^2$ pixels boxes, the minimal cluster size
for the experiment equals $2$ cells. With these technical adjustments,
we are able to reproduce the segregation indices and cell type ratios
reported by Méhes et al.~\cite{MehMonNemVic2012}, see
\figref{main_Florian_paper_plot_11_6} and
\figref{main_Florian_paper_plot_11_4}.

Note, that the time in the video is very coarsely labeled, as it is
limited to full hours. Since some hours included more frames than
others, we decided to linearly interpolate the time for each frame. For this, we
used the first frame with $1$h labeled and the first frame with $16$h
labeled, which is the latest full hour time stamp. This approximation together
with the limited resolution of the published video and the
cutoff from the original microscopy images, this explains the small
discrepancies between our data and the segregation indices reported in
Méhes et al.~\cite{MehMonNemVic2012}.

\subsection*{Cluster sizes}
We compute the distribution $\rho$ of cluster sizes for the models and the
experiment, see \figref{main_Florian_paper_plot_11_3} and
\figref{main_Florian_paper_plot_11_2}. Since the cellular automaton
and the video analysis of the experiment already yield a grid for each
time point with assigned cell types, only the output of the
Cahn-Hilliard model needs to be converted. Since the total area of
simulation for the cellular automaton and the Cahn-Hilliard model are
equal, due to our developed mapping, we project the Cahn-Hilliard grid
to the grid used by the cellular automaton. Each set of grid points of a
cluster has the same type and is connected to every other grid point in this set
through a sequence of von-Neumann neighborhoods within the cluster.
After identifying each cluster, we count the number of grid points
contained in each cluster and compute the corresponding area.

\subsection*{Pseuodo algorithm for the cellular automaton}

In order to simulate the cellular automaton in an effective way, we
implemented a version with continuous time by applying the idea of the
Gillespie-algorithm to the cellular automaton, instead of using the
usual Metropolis algorithm. This improved the performance of our
simulations drastically, in comparison to a simple algorithm with
discrete time steps and made the calibration to the experiments
feasible. The pseudo algorithm used for the cellular automaton reads
as follows:

\begin{enumerate}
  \item Initialise the lattice.
  \item Choose random one heterogeneous transition ($\textbf{x},\textbf{y}\in S, |\textbf{x}-\textbf{y}| = \Delta x \wedge \xi\neq\xi^{\textbf{x},\textbf{y}}$). Transitions with a higher rate $r(\textbf{x},\textbf{y})$, will be chosen with a linear higher probability $P(\xi\rightarrow\xi^{\textbf{x},\textbf{y}})$.
  \item The two cells of the selected transition will swap there position on the lattice.
  \item $\Delta t_\text{swap}$ is added to the time.
  \item If an end condition is reached, stop here.
  \item Else, return to step 2.
\end{enumerate}
\begin{equation}
  \begin{aligned}
    & P(\textbf{x},\textbf{y}) = \frac{r(\textbf{x},\textbf{y})}{\sum_{B}^{} r(e)}, e\in B, r(e) := r(\textbf{x},\textbf{y})\text{,}\\
    & B = \{\textbf{x},\textbf{y}\in S, |\textbf{x}-\textbf{y}| = \Delta x \wedge \xi\neq\xi^{\textbf{x},\textbf{y}}\}
  \end{aligned}\text{,}
\end{equation}
\begin{equation}
  \Delta t_\text{swap} \sim Exp(\lambda(\xi))\text{,}
\end{equation}
\begin{equation}
  \lambda(\xi) = \sum_{B}^{} \exp(\beta_\text{sum}(\textbf{x},\textbf{y},\xi))\text{.}
\end{equation}
Here $B$ denotes the set of all possible heterogeneity transitions.

\begin{figure}[ht!]
 \centering
 \includegraphics[scale=0.3]{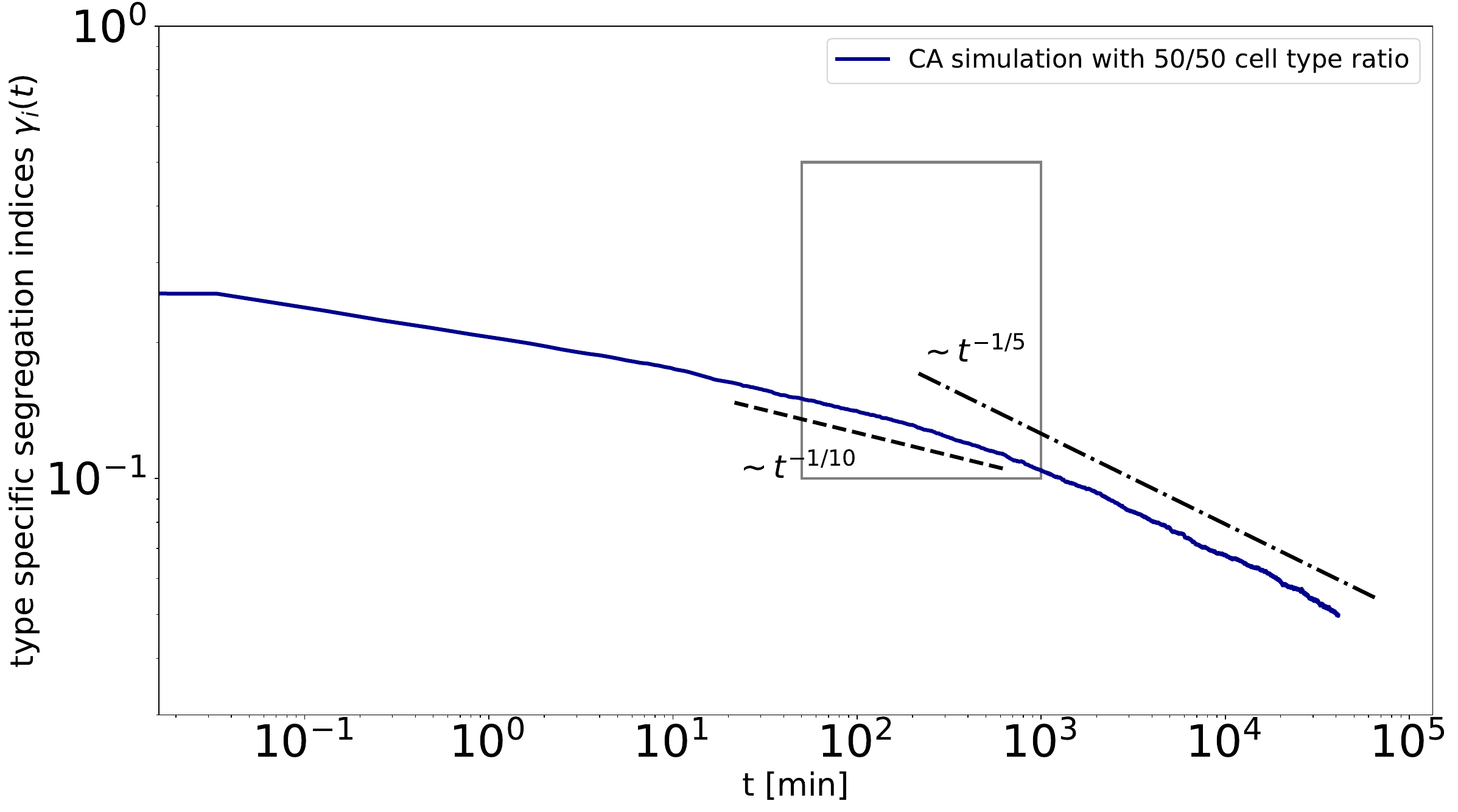}
 \caption{Test whether the scaling of the segregation index changes after
  the observations window. Even the cellular automaton simulation with a
  very flat initial progression does not reach the asymptotic scaling within
  the experimental regime, because the scaling still changes after the
  observation window (gray box) from $t^{-1/10}$ to $t^{-1/5}$. Shown are
  the segregation indices for a cellular automaton simulation with
  adhesion parameter $\textbf{$\beta$}=(0.44, -4.06, 0.44)^T$, time
  scale of migration $\tau=1\text{s}$, $140^2$ cells, and a cell type
  ratio of $50/50$. The flat early algebraic scaling with exponent $1/10$
  changes after the observation window of the experiment (gray box) to an
  algebraic scaling with exponent $1/5$. The scaling behavior
  is thus transitory. Note, that the chosen adhesion
  parameters translate to effective parameters $db=0$ and $b^*=9$ and
  correspond to the simulation with the lowest pseudo-algebraic scaling
  exponent in \figref{main_Florian_paper_plot_3_2}.}

 \label{fig:main_Florian_paper_plot_10_3}
\end{figure}

\begin{figure}[ht!]
 \centering
 \includegraphics[scale=0.3]{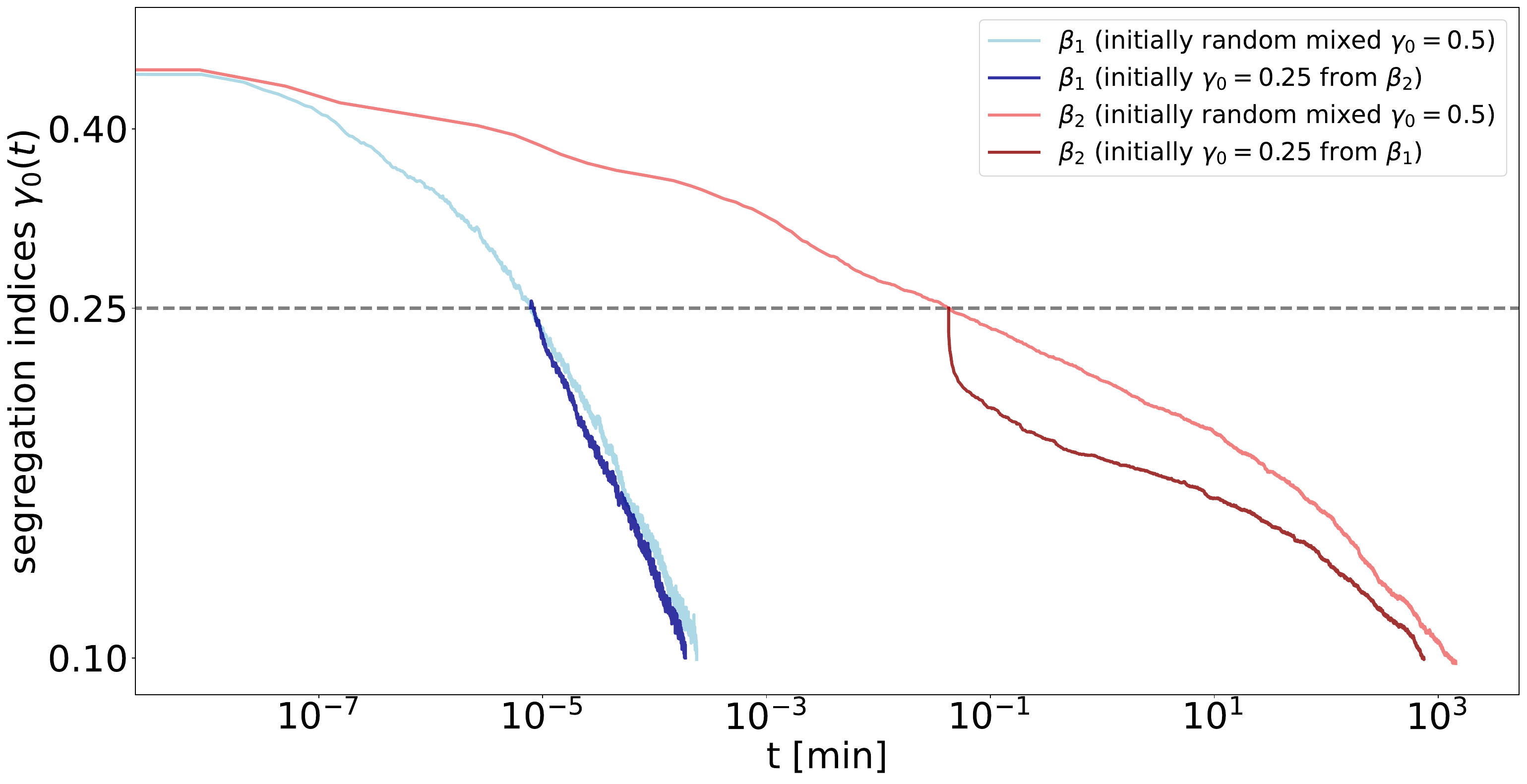}
 \caption{Illustration of the influence of initial conditions on the
   segregation process: Exemplary comparison of the segregation for
   different initial conditions in the cellular automaton. For each
   set of adhesion parameters $\text{\boldmath$\beta_1$} = (0.44,
   -4.06, 0.44)^T$ (blue lines), $\text{\boldmath$\beta_2$} =
   (-8.0,-5.5,0.0)^T$ (red lines) the simulation is started from a
   random initial condition $\gamma_0 \approx 0.5$ (brighter colored
   lines) and from a partially sorted field $\gamma_0 \approx 0.25$
   (darker colored lines), which resulted from segregation with the
   respective other set of adhesion parameters. All simulations used
   $100^2$ cells and a $50/50$ ratio (only $\gamma_0$
   shown). The parameters $\text{\boldmath$\beta_1$}$ are the
   $\gamma$-$\rho$-fitted adhesion parameters, see
   \figref{main_Florian_paper_plot_13_1}, the parameters
   $\text{\boldmath$\beta_2$}$ are from
   \figref{main_Florian_paper_plot_10_3}.}
 \label{fig:main_Florian_paper_plot_14_1}
\end{figure}

\begin{table}

  \begin{tabular}[h]{c|m{3cm}|m{4cm}|m{5cm}}
  \textbf{publication} & \textbf{environment} & \textbf{segregation index} & \textbf{average cluster diameter} \\
   & & \textbf{algebraic exponent} & \textbf{algebraic exponent} \\
  \hline
    Naso and Náraigh [26] & Cahn-Hilliard & $1/3$ (asymptotic) \\
    & & $\geq 1/6$ (transitory) \\
  \hline
    Naso and Náraigh [26] & Cahn-Hilliard & $2/3$ (asymptotic) \\
    Witkowski et al. [27] & Navier-Stokes & $\geq 1/6$ (transitory) \\
    Zhang et al. [28] &  &  \\
  \hline
    Glazier and Graner [12, 13] & CPM &
    logarithmic & \\
  \hline
    Nakajima and Ishihara [18] & CPM & $1/3$ (even mixtures) & $1/3$ (even mixtures) \\
     & & $1/4$ (uneven mixtures) & $1/4$ (uneven mixtures) \\
 \hline
   Durand [32] & CPM & $1/4$ \\
 \hline
   Cochet-Escartin et al. [36] & 3D CPM & $0.5$\\
 \hline
   Beatrici and Brunnet [19] & boids model & $0.18$ to $0.22$ or logarithmic \\
 \hline
   Strandkvist et al. [31] & boids model & $0.025$ to $0.17$ \\
  \hline
    Belmonte et al. [16] & self-propelled particle model with velocity alignment & $\leq 0.18$\\
  \hline
    Beatrici et al. [34] & active particle approach & & $1/4$ (without collective motion) \\
    & & & $1/2$ (collective motion) \\
  \hline
    Krajnc [35] & vertex model & $\leq 1/4$ \\
  \hline
    Krieg et al [8] & experiment & & $1/10$\\
  \hline
    Cochet-Escartin et al. [36] & 3D experiment & $0.74$ \\
  \hline
    Méhes et al. [20] & experiment & $0.31$ & $0.5$ to $0.74$ \\
  \end{tabular}
  \caption{Summary of the scalings previously published of studies used in the introduction.}
  \label{tab:scaling_summary}
\end{table}

\clearpage

\textbf{S1 Movie.} Supporting Information movie S1 provides cellular automaton exemplary segregation. (GIF)

\end{document}